\documentclass[11pt]{article}
\usepackage{amssymb}
\usepackage{amsfonts}
\usepackage{graphicx}
\usepackage{amsmath}
\usepackage{hyperref}

\makeatletter \@addtoreset{equation}{section} \makeatother

\newtheorem{theorem}{Theorem}[section]
\newtheorem{lemma}{Lemma}[section]

\newtheorem{remark}{Remark}[section]

\newtheorem{proposition}{Proposition}[section]

\newcommand{\mdet}{\mathrm{det}}
\newcommand{\intd}{\displaystyle\int}
\newcommand{\Tr}{\mathrm{Tr}\,}
\newcommand{\Str}{\mathrm{Str}\,}
\newcommand{\sdet}{\mathrm{sdet}\,}

\begin{document}

\title{Transfer matrix approach to 1d random band matrices: density of states}

\author{ Mariya Shcherbina
\thanks{Institute for Low Temperature Physics, Kharkiv, Ukraine, e-mail: shcherbina@ilt.kharkov.ua} \and
 Tatyana Shcherbina
\thanks{School of Mathematics, Institute for Advanced Study, Princeton, USA, e-mail: tshcherbina@ias.edu. Supported by NSF grant DMS-1128155.}}
\date{}
\maketitle

\begin{abstract}
We study the special case of
$n\times n$ 1D Gaussian Hermitian random band matrices,  when the covariance of the elements is determined by 
the matrix $J=(-W^2\triangle+1)^{-1}$. Assuming that $n\ge CW\log W\gg 1$, we prove that the averaged density of states coincides
with the Wigner semicircle law up to the correction of order $W^{-1}$.
\end{abstract}

\section{Introduction}\label{s:1}
We consider Hermitian $n\times n$ matrices $H_n$
  whose entries $H_{ij}$ are random
complex Gaussian variables with mean zero such that
\begin{equation}\label{ban}
\mathbf{E}\big\{ H_{ij}H_{lk}\big\}=\delta_{ik}\delta_{jl}J_{ij},
\end{equation}
where
\begin{equation}\label{J}
J_{ij}=\left(-W^2\Delta+1\right)^{-1}_{ij},
\end{equation}
and $\mathbf{E}\{\ldots\} $ denotes the average with respect to the probability distribution of $H_n$.
Here $\Delta$ is the discrete Laplacian on $\mathcal{L}=[1,n]\cap \mathbb{Z}$ with Neumann boundary conditions:
\begin{equation*}
(-\Delta f)_j=\left\{\begin{array}{ll}
f_1-f_2,& j=1;\\
2f_j -f_{j-1}-f_{j+1},& j=2,\ldots, n-1;\\
f_n-f_{n-1},& j=n.
 \end{array}
 \right.
\end{equation*}
 The probability law of  $H_n$ can be written in the form
\begin{equation}\label{band}
P_n(d H_n)=\prod\limits_{1\le i<j\le n}\dfrac{dH_{ij}d\overline{H}_{ij}}{2\pi J_{ij}}e^{-\frac{|H_{ij}|^2}{J_{ij}}}
\prod\limits_{i=1}^n\dfrac{dH_{ii}}{\sqrt{2\pi J_{ii}}}e^{-\frac{H_{ii}^2}{2J_{ii}}}.
\end{equation}
It is easy to see that  $J_{ij}\approx C_1W^{-1}\exp\{-C_2|i-j|/W\}$,  so it is exponentially small when $|i-j|\gg W$. Thus
matrices $H_n$ can be considered as a special case of random band matrices with the band width $W$.  The same model can be defined similarly in any dimension $d$
 (then $i,j\in\mathcal{L}=[1,n]^d\cap \mathbb{Z}^d$).

  Let $\lambda_1^{(n)},\ldots,\lambda_n^{(n)}$ be the eigenvalues of
$H_n$. Define their Normalized Counting Measure
(NCM) as
\begin{equation*}
\mathcal{N}_n(I)=\dfrac{1}{n}\sharp\{\lambda_j^{(n)}\in
I,j=1,\ldots,n \},\quad \mathcal{N}_n(\mathbb{R})=1,
\end{equation*}
where $I$ is an arbitrary interval of the real axis.
It was shown in \cite{BMP:91, MPK:92} that for 1d RBM (even for more general than (\ref{J}) form of the variance) $\mathcal{N}_n$ converges  in probability, as $n,W\to\infty$, to a non-random measure
$\mathcal{\mathcal{N}}$, which  is absolutely continuous,
and its density $\rho$ is given by the well-known Wigner semicircle law (the same result
is valid for Wigner ensembles, in particular, for Gaussian ensembles GUE, GOE):
\begin{equation}\label{rho_sc}
\rho_{sc}(E)=\left\{\begin{array}{cc}
(2\pi)^{-1}\sqrt{4-E^2},& E\in[-2,2];\\
0,& |E|\ge 2.
\end{array}
\right.
\end{equation}
A substantial interest to random band matrices is caused by the fact that they are natural intermediate models between random
Schr$\ddot{\hbox{o}}$dinger matrices $ H_{RS}=-\Delta+\lambda V$, in which the randomness
only appears in the diagonal potential $V$ ($\lambda$ is a small parameter which
measures the strength of the disorder) and
mean-field random matrices such as $n\times n$ Wigner matrices, i.e. Hermitian
random matrices with i.i.d elements. In particular, RBM can be used to model the Anderson metal-insulator phase transition. 
Moreover, it is conjectured  (see \cite{Ca-Co:90, FM:91}) that the transition for RBM can be investigated even in $d=1$ by varying the band width $W$. 
It is expected that 1d RBM changes the spectral local behaviour of random operator type with
Poisson local eigenvalue statistics corresponding to localized eigenstates (for $W\ll \sqrt{n}$) to the local
spectral behaviour of the Gaussian Unitary Matrix type corresponding to delocalized eigenstates (for $W\gg \sqrt{n}$)
(for more details on these conjectures see e.g. \cite{Sp:12}).
Some partial results about  localization and delocalization (in a weak sense) for  general RBM  was obtained 
in \cite{S:09}, \cite{EK:11}, \cite{Yau:12}. Universality of the gap distribution for $W\sim n$ was also obtained
in a recent paper \cite{BEYY:16}. However,  the question of the existing of a crossover in RBM
is still open even for $d=1$.

One of the  approaches, which allows to work with random operators with non-trivial spatial structures, is supersymmetry techniques (SUSY) based
on the representation of the determinant as an integral over the Grassmann variables. 
This method is widely  used in the physics literature  and is potentially very powerful, but the rigorous
control of the integral representations, which can be obtained by this method, is quite difficult.
However, it can be done rigorously for some special class of RBM.  For instance, by using SUSY
the detailed information about the averaged density of states of ensemble (\ref{ban}) -- (\ref{band}) in dimension 3 including local semicircle low
at arbitrary short scales  and smoothness in 
energy (in the limit of infinite volume and fixed large band width $W$) was obtained in \cite{DPS:02}. Moreover,
by applying SUSY approach in  \cite{TSh:14}, \cite{SS:16} the crossover in this model (in 1d) was proved for the correlation functions
of characteristic polynomials. In addition, the rigorous application of SUSY to the Gaussian RBM  which has the special block-band structure 
(special case of Wegner's orbital model)
was developed in \cite{TSh:14_1}, where the universality of the bulk local regime for $W\sim n$ was proved.
Combining this approach with Green's function comparison strategy the delocalization (in a strong sense) for $W\gg n^{6/7}$  has been proved  in \cite{EB:15} for the block band matrices with rather general element's distribution.

In this paper we develop the method of \cite{SS:16}, which combines the SUSY techniques with a transfer matrix approach (see also \cite{D-S:15}).
The final goal is to extend this method from the correlation function of characteristic polynomials  to usual 
correlation functions of (\ref{ban}) -- (\ref{band}), which allows to study the crossover of local eigenvalue statistics for 1d RBM. 
To this end we have to study the transfer operator involving not only the complex, but also the  Grassmann variables 
(for the second correlation functions it involves 8
Grassmann variables). At the present paper we make the first step in this direction: we study the transfer operator appearing from the integral representation of the density of states of ensemble (\ref{ban}) -- (\ref{band}) in 1d (see (\ref{fin}) below), which has only two Grassmann variables.
The supersymmetric transfer matrix formalism was first suggested by Efetov (see \cite{Ef}), and it was successfully  applied rigorously 
to the density of states of some models
(see e.g. \cite{CK:86}, \cite{C:88}).

According to the property of the Stieltjes transform, the averaged density of states is given by
\begin{equation}\label{den}
\bar \rho_n(E)=\dfrac{1}{\pi}\lim\limits_{\varepsilon\downarrow 0}\mathbf{E}\Big\{n^{-1} \hbox{Im}\,\Tr (E_\varepsilon-H_n)^{-1}\Big\},
\end{equation}
where $E_\varepsilon=E-i\varepsilon$, $E\in (-2,2)$.

Thus, we are interested in 
\begin{equation}\label{g_E}
\bar g_n(E)=\lim\limits_{\varepsilon\downarrow 0} \bar g_n(E_\varepsilon)
\end{equation}
with
\begin{equation*}
\bar g_n(E_\varepsilon)=\mathbf{E}\Big\{n^{-1}\Tr (E_\varepsilon-H_n)^{-1}\Big\}=\mathbf{E}\Bigg\{-\dfrac{\partial}{\partial x}\dfrac{\mdet (E_\varepsilon-H_n)}
{\mdet (E_\varepsilon+x/n-H_n)}\Bigg|_{x=0}\Bigg\}.
\end{equation*}
Our main result is
\begin{theorem}\label{t:0} 
Let $H_n$ be 1d Gaussian RBM defined in (\ref{ban}) -- (\ref{band}) with $n\ge C_0W\log W$, and let $|E|\le 4\sqrt{2}/3\approx 1.88$. Then
for $\bar g_n(E)$ defined in (\ref{g_E}) we have
\begin{equation*}
|\bar g_n(E)-g_{sc}(E)|\le C/W,
\end{equation*}
where 
\[
g_{sc}(E)=\lim\limits_{\varepsilon\downarrow 0}\int\dfrac{\rho_{sc}(\lambda)d\lambda}{E-i\varepsilon-\lambda}=\dfrac{E+i\sqrt{4-E^2}}{2}.
\]
In particular,
\begin{equation*}
|\bar \rho_n(E)-\rho_{sc}(E)|\le C/W,
\end{equation*}
where $\bar \rho_n(E)$ is an averaged density of states (\ref{den}), and $\rho_{sc}$ is defined in (\ref{rho_sc}).
\end{theorem}
Note that Theorem \ref{t:0} gives
\begin{equation}\label{loc_sc}
|\bar g_n(E-i\varepsilon)-g_{sc}(E-i\varepsilon)|\le C/W
\end{equation}
uniformly in any arbitrary small $\varepsilon\ge 0$. As it was mentioned above, similar asymptotics (with correction $C/W^2$) for RBM of (\ref{ban}) in 3d
was obtained in \cite{DPS:02}, however their method cannot be applied to 1d case. All other previous results about the density of states
for RBM deal with $\varepsilon\gg~W^{-1}$ or bigger (for fixed $\varepsilon>0$ the asymptotics  (\ref{loc_sc}) follows from the results of \cite{BMP:91};
\cite{EK:11} gives (\ref{loc_sc}) with $\varepsilon\gg W^{-1/3}$; \cite{S:11} yields (\ref{loc_sc}) for 1d RBM with Bernoulli elements distribution for
$\varepsilon\ge W^{-0.99}$, and \cite{EYY:10} proves similar to (\ref{loc_sc})  asymptotics with correction $1/(W\varepsilon)^{1/2}$ for $\varepsilon\gg 1/W$).
On the other hand, the methods of \cite{EK:11}, \cite{EYY:10} allow to control $n^{-1}\Tr (E_\varepsilon-H_n)^{-1}$  and $(E_\varepsilon-H_n)^{-1}_{xy}$ 
for $\varepsilon\gg W^{-1}$ without expectation, which gives some information about the localization length. This cannot be obtained from Theorem \ref{t:0}, since it requires
estimates on $\mathbb{E}\{|(E_\varepsilon-H_n)^{-1}_{xy}|^2\}$. Similar estimates for $\varepsilon\approx n^{-1}$ is required to work with 
the second correlation function, and we hope it  will be the aim of the next paper. 

The paper is organized as follows. In Section \ref{s:2} we re-derive an integral representation for $\bar g_n(E)$ obtained in \cite{DPS:02}.  
In Section \ref{s:3} we rewrite 
this representation in terms of the transfer operator $\mathcal{K}$ (see (\ref{main1})). Section \ref{s:4} deals with the analysis of the operator $\mathcal{K}$ (see Theorem \ref{t:1}) 
and the proof of Theorem
\ref{t:0}. In Section \ref{s:5} we prove an important preliminary result needed for Section \ref{s:4}. 

\section{Integral representation}\label{s:2}
In this section we obtain an integral representation for $\bar g_n(E)$ of (\ref{g_E}) by using integration
over the Grassmann variables. Such representation for the density of states of ensemble (\ref{ban}) -- (\ref{J}) was
obtained in \cite{DPS:02} in any dimension $d$. For the reader convenience we repeat here the derivation of
the integral representation for $d=1$. 

Integration over the Grassmann variables has been introduced by Berezin and is widely used in the physics
literature (see e.g. \cite{Ef}). A brief outline of the techniques can be found e.g. in  \cite{Ef}.

Let $A$ be an ordinary matrix with a positive Hermitian part. The following Gaussian
integral is well-known:
\begin{equation}\label{G_C}
\intd \exp\Big\{-\sum\limits_{j,k=1}^nA_{j,k}z_j\overline{z}_k\Big\} \prod\limits_{j=1}^n\dfrac{d\,\Re
z_jd\,\Im z_j}{\pi}=\dfrac{1}{\mdet A}.
\end{equation}
One of the most important formulas of the Grassmann variables theory is the analog of (\ref{G_C}) for the
Grassmann variables (see \cite{Ef}):
\begin{equation}\label{G_Gr}
\int \exp\Big\{-\sum\limits_{j,k=1}^nA_{j,k}\overline{\psi}_j\psi_k\Big\}
\prod\limits_{j=1}^nd\,\bar\psi_jd\,\psi_j=\mdet A,
\end{equation}
where $A$ now is any $n\times n$ matrix.
Combining these two formulas one can obtain also
\begin{equation}\label{G_sup}
\intd \exp\Big\{-\Phi^+F\Phi\Big\} \prod\limits_{j=1}^n\dfrac{d\,\Re
z_jd\,\Im z_j\,d\bar\psi_j\,d\psi_j}{\pi}=\sdet F,
\end{equation}
where $\Phi=(z_1,\ldots, z_n,\psi_1,\ldots,\psi_n)^t$,
\begin{equation*}
F=\left(\begin{array}{cc}
B&\Sigma^+\\
\Sigma &A
\end{array}\right),\quad \sdet F=\dfrac{\mdet(A-\Sigma B^{-1}\Sigma^+)}{\mdet B},
\end{equation*}
and $B>0$, $A$ are $n\times n$ complex matrices, $\Sigma, \Sigma^+$ are $n\times n$ matrices of anticommuting elements of Grassmann algebra.

Using (\ref{G_C}) -- (\ref{G_Gr}), we can rewrite
\begin{align*}\notag
&\mathbf{E}\Big\{\dfrac{\mdet (E_\varepsilon-H_n)}{\mdet (E_\varepsilon+x/n-H_n)}\Big\}=\mathbf{E}\bigg\{\displaystyle\int
e^{i\sum\limits_{j,k=1}^n(E_\varepsilon-H_n)_{jk}\overline{\psi}_{j}\psi_{k}+i\sum\limits_{j,k=1}^n(E_\varepsilon+x/n-H_n)_{jk}\overline{\phi}_{j}\phi_{k}}
d\Phi\bigg\}\\
&=\displaystyle\int \exp\Big\{\sum_{j}\Big(iE_\varepsilon\overline{\psi}_{j}\psi_{j}+i(E_\varepsilon+x/n)\overline{\phi}_{j}\phi_{j}\Big)\Big\}\\ \notag
&\times \mathbf{E}\bigg\{\exp\Big\{-\sum\limits_{j<k}\Big(i \Re
H_{jk}\cdot(\eta_{jk}+\eta_{kj})
-\Im H_{jk}\cdot(\eta_{jk}-\eta_{kj})\Big)-i\sum_{j}H_{jj}\cdot \eta_{jj}\Big\}
\bigg\}d\Phi,
\end{align*}
where $\{\psi_{j}\}_{j=1}^n$ are Grassmann (i.e. anticommuting) variables, $\{\phi_{j}\}_{j=1}^n\in \mathbb{C}^n$,
\begin{align*}
&\eta_{jk}=\bar\psi_j\psi_k+\bar\phi_j\phi_k,\\
&d\Phi=\prod\limits_{q=1}^n\dfrac{d\Re \phi_q\, d\Im \phi_q}{\pi}\prod\limits_{q=1}^n
d\,\bar\psi_{q}d\,\psi_{q}.
\end{align*}
Taking the average according to (\ref{band}), we get
\begin{align}\label{av}
&\mathbf{E}\Big\{\dfrac{\mdet (E_\varepsilon-H_n)}{\mdet (E_\varepsilon+x/n-H_n)}\Big\}= 
\displaystyle\int e^{\sum_{j}\big(iE_\varepsilon\overline{\psi}_{j}\psi_{j}+i(E_\varepsilon+x/n)\overline{\phi}_{j}\phi_{j}\big)
-\frac{1}{2}\sum\limits_{j,k}J_{jk}\eta_{jk}\eta_{kj}}d\Phi\\ \notag
&=\displaystyle\int d\Phi\,\, \exp\Big\{\sum_{j}\big(iE_\varepsilon\overline{\psi}_{j}\psi_{j}+i(E_\varepsilon+x/n)\overline{\phi}_{j}\phi_{j}\big)\Big\}\\ \notag
&\exp\Big\{\frac{1}{2}\sum\limits_{j,k}J_{jk}\bar\psi_j\psi_j\cdot \bar\psi_k\psi_k-\sum\limits_{j,k}J_{jk}\bar\psi_j\phi_j\cdot \psi_k\bar\phi_k
-\frac{1}{2}\sum\limits_{j,k}J_{jk}\bar\phi_j\phi_j\cdot \bar\phi_k\phi_k\Big\}.
\end{align}
To convert the quartic interaction in (\ref{av}) into a quadratic one we perform a standard  Hubbard-Stratonovich
transformation:
\begin{align*}
&\exp\Big\{\frac{1}{2}\sum\limits_{j,k}J_{jk}\bar\psi_j\psi_j\cdot \bar\psi_k\psi_k\Big\}=\dfrac{\mdet^{-1/2} J}{(2\pi)^{n/2}}\int \exp\{-\frac{1}{2}\sum\limits_{j,k}J^{-1}_{jk}a_ja_k+\sum_j a_j\bar\psi_j\psi_j\}
\prod\limits_{j=1}^nda_j;\\ \notag
&\exp\Big\{-\frac{1}{2}\sum\limits_{j,k}J_{jk}\bar\phi_j\phi_j\cdot \bar\phi_k\phi_k\Big\}=\dfrac{\mdet^{-1/2} J}{(2\pi)^{n/2}}\int \exp\{-\frac{1}{2}\sum\limits_{j,k}J^{-1}_{jk}b_jb_k-i\sum_j b_j\bar\phi_j\phi_j\}
\prod\limits_{j=1}^ndb_j;
\end{align*}
\begin{multline*}
\exp\Big\{-\sum\limits_{j,k}J_{jk}\bar\psi_j\phi_j\cdot\psi_k \bar\phi_k\Big\}\\
=\mdet\, J\int \exp\{-\sum\limits_{j,k}
J^{-1}_{jk}\bar\rho_j\rho_k-i\sum_j \bar\rho_j\psi_j\bar\phi_j+i\sum_j \rho_j\bar\psi_j\phi_j\}
\prod\limits_{j=1}^nd\bar\rho_j\,d\rho_j.
\end{multline*}
This gives
\begin{align}\label{str1}
&\mathbf{E}\Big\{\dfrac{\mdet (E_\varepsilon-H_n)}{\mdet (E_\varepsilon+x/n-H_n)}\Big\}
=\dfrac{1}{(2\pi)^n}\displaystyle\int d\Phi\,d \bar X \,\, e^{\sum_{j}\big(iE_\varepsilon\overline{\psi}_{j}\psi_{j}+i(E_\varepsilon+x/n)\overline{\phi}_{j}\phi_{j}\big)}
\\ \notag
&\exp\Big\{-\dfrac{1}{2}\sum\limits_{j,k}J^{-1}_{jk}\,\Str X_jX_k+\sum\limits_{j}a_j\bar\psi_j\psi_j-i\sum_j b_j\bar\phi_j\phi_j-i\sum_j \bar\rho_j\psi_j\bar\phi_j+i\sum_j \rho_j\bar\psi_j\phi_j\Big\},
\end{align}
where
\begin{equation}\label{x_j}
X_j=\left(\begin{array}{cc}
b_j&\bar\rho_j\\
\rho_j&ia_j
\end{array}\right),\quad \Str \left(\begin{array}{cc}
x&\bar\sigma\\
\sigma&y
\end{array}\right)=x-y,
\end{equation}
and
\[
d\bar X=\prod\limits_{j=1}^nd X_j,\quad dX_j=da_j\,db_j\,d\bar\rho_j\,d\rho_j.
\]
Here $a_j$, $b_j$ are complex variables, and $\bar\rho_j$, $\rho_j$ are Grassmann variables. 
Applying (\ref{G_sup}) to integrate over $d\Phi$ in (\ref{str1}), we obtain
\begin{multline}\label{str2}
\mathbf{E}\Big\{\dfrac{\mdet (E_\varepsilon-H_n)}{\mdet (E_\varepsilon+x/n-H_n)}\Big\}\\
=\dfrac{1}{(2\pi)^n}\displaystyle\int \exp\Big\{-\dfrac{1}{2}\sum\limits_{j,k}J^{-1}_{jk}\Str X_jX_k\Big\}
\prod\limits_{j=1}^n\sdet (X_j-\Lambda_x)\,d\bar X ,
\end{multline}
where 
\begin{equation*}
 \Lambda_x=\left(\begin{array}{cc}
E_\varepsilon+x/n&0\\
0& E_\varepsilon
\end{array}\right).
\end{equation*}
Substituting (\ref{J}) and (\ref{x_j}), we can rewrite (\ref{str2}) as 
\begin{align}\notag
&\lim\limits_{\varepsilon\downarrow 0}\mathbf{E}\Big\{\dfrac{\mdet (E_\varepsilon-H_n)}{\mdet (E_\varepsilon+x/n-H_n)}\Big\}
=\dfrac{1}{(2\pi)^n}\displaystyle\int \exp\Big\{-\dfrac{W^2}{2}\sum\limits_{j=1}^n\Big((a_j-a_{j-1})^2+(b_j-b_{j-1})^2\Big)\Big\}\\ \label{fin1}
&\exp\{-W^2\sum\limits_{j=1}^n(\bar\rho_j-\bar\rho_{j-1})(\rho_j-\rho_{j-1})-\dfrac{1}{2}\sum\limits_{j=1}^n(a_j^2+b_j^2)-\sum\limits_{j=1}^n\bar\rho_j\rho_j\Big\}\\
\notag & \prod\limits_{j=1}^n\dfrac{ia_j-E_\varepsilon}{b_j-E_\varepsilon-x/n} \cdot \prod\limits_{j=1}^n\Big(1+\dfrac{\bar\rho_j\rho_j}{(ia_j-E_\varepsilon)(b_j-E_\varepsilon-x/n)}\Big)\,d\bar X\\
\notag &=\dfrac{1}{(2\pi)^n}\displaystyle\int \, d\bar X\, \exp\Big\{-\dfrac{W^2}{2}\sum\limits_{j=1}^n\Big((a_j-a_{j-1})^2+(b_j-b_{j-1})^2\Big)\Big\}\\
\notag&\times \exp\Big\{-\dfrac{1}{n}\sum\limits_{j=1}^n ((b_j+i
\sqrt{4-E^2}/2)x+x^2/2n)-\sum\limits_{j=1}^n(f_a(a_j)+f_b(b_j)) \Big\}\\
\notag & \times \exp\{-W^2\sum\limits_{j=1}^n(\bar\rho_j-\bar\rho_{j-1})(\rho_j-\rho_{j-1})-\sum\limits_{j=1}^n\bar\rho_j\rho_j L(a_j,b_j)\Big\},
\end{align}
where, to obtain the last equality, we use 
\[
1+\dfrac{\bar\rho_j\rho_j}{(ia_j-E_\varepsilon)(b_j-E_\varepsilon-x/n)}=\exp\Big\{\dfrac{\bar\rho_j\rho_j}{(ia_j-E_\varepsilon)(b_j-E_\varepsilon-x/n)}\Big\},
\]
change $b_j\to b_j+x/n+\dfrac{i\sqrt{4-E^2}}{2}$, $a_j\to a_j-iE_\varepsilon/2$ and put $\varepsilon=0$. In the last line of (\ref{fin1}) we put
\begin{align}\label{f_a,b}
&f_a(x)= (x-iE/2)^2/2 -\log (ix-E/2)+C^*;\\
& f_b(x)= (x+i\sqrt{4-E^2}/2)^2/2 +\log (x-E+i\sqrt{4-E^2}/2)-C^*;\notag\\ \notag
&C^*=(E/2+i\sqrt{4-E^2}/2)^2/2 +\log (-E/2+i\sqrt{4-E^2}/2);\\ 
&L(x,y)=1-\dfrac{1}{(ix-E/2)(y-E+i
\sqrt{4-E^2}/2)}.
\label{L}\end{align}
Taking the derivative of (\ref{fin1}) with respect to $x$ and putting $x=0$, we get
\begin{align}\notag
&\bar g_n(E)=\dfrac{1}{(2\pi)^n}\displaystyle\int \Big(\sum\limits_{j=1}^n \dfrac{b_j+i\sqrt{4-E^2}/2}{n}\Big)\cdot \exp\Big\{-\dfrac{W^2}{2}\sum\limits_{j=1}^n\Big((a_j-a_{j-1})^2+(b_j-b_{j-1})^2\Big) \Big\}\\ \label{fin}
 & \exp\{-W^2\sum\limits_{j=1}^n(\bar\rho_j-\bar\rho_{j-1})(\rho_j-\rho_{j-1})-\sum\limits_{j=1}^n(f_a(a_j)+f_b(b_j))-\sum\limits_{j=1}^n\bar\rho_j\rho_j L(a_j,b_j)\Big\}\, d\bar X.
\end{align}
Besides, putting $x=0$ (\ref{fin1}) we have
\begin{align}\notag
&1=\dfrac{1}{(2\pi)^n}\displaystyle\int  \exp\Big\{-\dfrac{W^2}{2}\sum\limits_{j=1}^n\Big((a_j-a_{j-1})^2+(b_j-b_{j-1})^2\Big) \Big\}\\ \label{1_rep}
 & \exp\{-W^2\sum\limits_{j=1}^n(\bar\rho_j-\bar\rho_{j-1})(\rho_j-\rho_{j-1})-\sum\limits_{j=1}^n(f_a(a_j)+f_b(b_j))-\sum\limits_{j=1}^n\bar\rho_j\rho_j L(a_j,b_j)\Big\}\, d\bar X.
\end{align}
Now let us study the stationary points of $f_a$, $f_b$:
\begin{lemma}\label{l:f_a,b}
\begin{description}
\item{(i)\,\,} The function $\Re f_a(x)$, $x\in \mathbb{R}$ attains its minimum at 
\begin{align}\label{a_pm}
a_\pm=\pm\dfrac{\sqrt{4-E^2}}{2}.
\end{align}
Moreover,
\[
f_a(a_+)=f_a'(a_{\pm})=0, \quad f_a(a_-)=\dfrac{iE\sqrt{4-E^2}}{2}+2\log(-E/2+i\sqrt{4-E^2/2})\in i\mathbb{R}.
\]
\item{(ii)\,\,} For $|E|< 4\sqrt 2/3\approx 1.88$, the function $\Re f_b(x)$, $x\in \mathbb{R}$ attains its minimum at  
\begin{align}\label{b_s} b_s=\dfrac{E}{2}.
\end{align}
Moreover,
\[
f_b(b_s)=f_b'(b_s)=0.
\]
\end{description}
\end{lemma}
The proof of the lemma is straightforward and it is omitted here.

Define also
\begin{align}\label{c_pm}
c_\pm:=f''_a(a_\pm)/2,
\end{align}
and note that $\Re c_\pm > 0$, $\arg c_\pm\in (-\pi/2,\pi/2)$. Besides,
\begin{align}\label{L_pm}
&f''_b(b_s)=2c_+;\\ \notag
&L^+:=L(a_+,b_s)=1-\Big(\dfrac{E-i\sqrt{4-E^2}}{2}\Big)^2=2c_+;\\
&L^-:=L(a_-,b_s)=0.
\notag\end{align}

\section{Transfer matrix approach}\label{s:3}
Expanding the exponent into the series it is easy to see that
\begin{align*}
&\int\exp\Big\{-W^2(\bar\rho'-\bar\rho)(\rho'-\rho)-L\bar \rho'\rho'\Big\} \Big(q_1+q_2\bar\rho'\rho'+q_3\bar\rho'+q_4\rho'\Big)d\rho'\, d\bar\rho'\\
&=W^2\Big(((1+L/W^2)q_1-q_2/W^2)+(q_2-Lq_1)\bar\rho\rho +q_3\bar\rho+q_4\rho\Big),
\end{align*}
which means that the Grassmann part of the operator acts on a vector $q=(q_1,q_2,q_3,q_4)\in \mathbb{C}^4$ as $4\times 4$ matrix $W^2 \mathcal{Q}$ with a  block matrix
\begin{align*}
\mathcal{Q}=\left(\begin{array}{cc}\breve{Q}&0\\0&I\end{array}\right),
\end{align*}
where the $2\times 2$ matrix $\breve{Q}$ has the form
\begin{align*}
\breve{Q}=\left(\begin{array}{cc}
1+L/W^2&-1/W^2\\
-L&1
\end{array}\right)
\end{align*}
with the function $L$ of (\ref{L}). 

Note that the usual $\mathbb{C}^4$ inner product of vectors $q=\{q_i\}_{i=1}^4$ and $p=\{p_i\}_{i=1}^4$ does not coincide with the product
of two corresponding Grassmann polynomials $q_1+q_2\bar\rho\rho+q_3\bar\rho+q_4\rho$, $p_1+p_2\bar\rho\rho+p_3\bar\rho+p_4\rho$.
For instance, the product of  $q_1+q_2\bar\rho\rho$, $p_1+p_2\bar\rho\rho$ is obtained by the usual inner product of
$(q_1, q_2,0,0)$ and $(p_2, p_1, 0, 0)$ (not $(p_1, p_2, 0, 0)$).

Introduce compact integral operators $A$ and $A_1$ in $L_2[\mathrm{R}]$ with the kernels
\begin{align}
\label{A}
&\,A(a_1,a_2)=\mathcal{F}_0(a_1)B(a_1,a_2)\mathcal{F}_0(a_2), \quad \mathcal{F}_0(a)=e^{-f_a(a)/2};\\
 \label{A_1}
&A_1(b_1,b_2)=\mathcal{F}_1(b_1)B(b_1,b_2)\mathcal{F}_1(b_2), \quad\,\, \,\mathcal{F}_1(b)=e^{-f_b(b)/2};\\
&\,B(a_1,a_2)=(2\pi)^{-1/2}We^{-W^2(a_1-a_2)^2/2}, 
\notag
\end{align}
where $f_a$ and $f_b$ are defined in (\ref{f_a,b}). 
Denote also
\begin{align}\label{K}
K=A\otimes A_1,\quad \breve{K}=\left(\begin{array}{cc}K&0\\0&K\end{array}\right),\quad
\breve{K}_1=\left(\begin{array}{cc}\breve K&0\\0&\breve K\end{array}\right).
\end{align}
Let  $\mathcal{B}$ be the operator of multiplication by 
\begin{align*}
d(b)=b+i\sqrt{4-E^2}/2.
\end{align*}
Set
\begin{align}\label{br_B}
&\breve{\mathcal{B}}=\left(\begin{array}{cc}\mathcal{B}&0\\0&\mathcal{B}\end{array}\right),\quad
\breve{\mathcal{B}}_1=\left(\begin{array}{cc}\breve{\mathcal{B}}&0\\0&\breve{\mathcal{B}}\end{array}\right),\\
&e_1'=(1,0,0,0)^t,\quad e'(L)=(-\bar L,1,0,0)^t,\notag\\ 
&e_1=(1,0)^t,\quad e_2=(0,1)^t,\quad e(L)=(-\bar L,1)^t, \quad e_L=(1,-L/W)^t.
\notag\end{align}
With these notations we can rewrite (\ref{fin}) as
\begin{align}\label{main}
\bar g_n(E)=&\frac{1}{n}\sum_{j=0}^{n-1}\Big( \big((\breve K_1\mathcal{Q})^j\breve{\mathcal{B}}_1(\breve K_1\mathcal{Q})^{n-1-j}e_1',e'(L)\big)_4
\mathcal{F},\bar{\mathcal{F}}\Big)\\
=&\frac{1}{n}\sum_{j=0}^{n-1}\Big( \big((\breve K\breve{Q})^j\breve{\mathcal{B}}(\breve K\breve{Q})^{n-1-j}e_1,e(L)\big)_2
\mathcal{F},\bar{\mathcal{F}}\Big)
\notag\end{align}
where $(.,.)_4$ and $(.,.)_2$ mean  inner products in $\mathbb{C}^4$ and $\mathbb{C}^2$ respectively,
\[\mathcal{F}(a,b)=\mathcal{F}_0(a)\mathcal{F}_1(b),\]
and we have used the block-diagonal structure of $\breve K_1$,  $\breve{\mathcal{B}}_1$, and $\mathcal{Q}$.

To study the r.h.s. of (\ref{main}), let us
rewrite it  in a more convenient form. Introduce the matrices
\begin{align}\label{T_W}
T=\left(\begin{array}{cc}0&W^{-1/2}\\W^{1/2}&0\end{array}\right),\quad S=\left(\begin{array}{cc}1&-L/W\\-1/W&1+L/W^2\end{array}\right).
\end{align}
Note that for $|E|>\varepsilon_*$ with any fixed $\varepsilon_*>0$
\begin{equation}\label{b_L}
|L(a_j,b_j)|<L_0,\quad a_j\in\mathbb{R},\quad  b_j\in\mathbb{R},
\end{equation}
 for some constant $L_0$ depending on $E$, and
hence  
\begin{equation}\label{norm_S}
\|S\|\le 1+C_0/W,
\end{equation}
where one can take $C_0=2+L_0$.
\begin{remark}\label{r:b_L}
In order to have the bound (\ref{b_L}) valid for any $E$, in the case of $|E|\le\varepsilon_*$ with sufficiently small  $\varepsilon_*>0$ we
deform the contour of integration with respect to $a$ in some small neighbourhood $U_{\tilde\varepsilon}$ of $a=0$ in such a way that guarantees 
the conditions
\begin{align}\label{b_L.1}
\sup_{a\in U_{\tilde\varepsilon}}\{|\cos^{-1/2}2\phi(a)|\}\sup_{a\in U_{\tilde\varepsilon}}\{e^{-\Re f_a(a)/2}\}\le 1-\delta,
\end{align}
where $\phi(a)$ is an angle between the contour and the real line at the point $a$ and $\delta>0$ is some fixed number which can be chosen
as small as we want  (see Section 4). The above condition will be important below (see the proof of (\ref{u_2})).
\end{remark}

It is easy to check that
\[\breve Q=TST^{-1}.\]
Hence, since
\[T^{-1}\breve KT=\breve K,\quad T^{-1}\breve{\mathcal{B}} T=\breve{\mathcal{B}},\]
we have
\begin{align*}\notag
\Big( \big((\breve K\breve{Q})^j\breve{\mathcal{B}}(\breve K\breve{Q})^{n-1-j}e_1,e(L)\big)_2
\mathcal{F},\bar{\mathcal{F}}\Big)=&\Big( \big((\breve KS)^j\breve{\mathcal{B}}(\breve KS)^{n-1-j}T^{-1}e_1,T^*e(L)\big)_2
\mathcal{F},\bar{\mathcal{F}}\Big)\\
=&W\Big( \big((\breve KS)^j\breve{\mathcal{B}}(\breve KS)^{n-1-j}e_2,e_L\big)_2
\mathcal{F},\bar{\mathcal{F}}\Big)
\end{align*}
since
\[T^{-1}e_1=W^{1/2}(0,1)^t=W^{1/2}e_2,\quad T^*e(L)=W^{1/2}(1,-L/W)^t=W^{1/2}e_L\]
Thus, denoting
\begin{align}\label{cal_K}
\mathcal{K}=\breve KS
\end{align}
we get finally from (\ref{main})
\begin{align}\label{main1}
\bar g_n(E)=&\frac{W}{n}\sum_{j=0}^{n-1}\Big( \big(\mathcal{K}^j\breve{\mathcal{B}}\mathcal{K}^{n-1-j}e_2,e_L\big)_2
\mathcal{F},\bar{\mathcal{F}}\Big),
\end{align}
In what follows it will be important for us that the same argument applied to (\ref{1_rep}) implies
\begin{align}\label{main2}
1=W\Big( \big(\mathcal{K}^{n-1}e_2,e_L\big)_2
\mathcal{F},\bar{\mathcal{F}}\Big).
\end{align}

\section{Analysis of   $\mathcal{K}$}\label{s:4}
In this section we apply  the method developed in \cite{SS:16}, based on
 the  proposition, which is a standard  linear algebra tool 
 \begin{proposition}\label{p:sp}
Given a compact operator $\mathcal{K}$, assume that there is an orthonormal basis $\{\Psi_l\}_{l\ge 0}$ such that
the resolvent
\[\widehat{\mathcal{G}}_{jk}(z)=(\widehat{\mathcal{K}}-z)^{-1}_{jk},\quad
\widehat{\mathcal{K}}=\{\mathcal{K}_{jk}\}_{j,k=1}^{\infty}\]
is uniformly bounded in   $z\in\Omega\subset \mathbb{C}$, where $\Omega$ is some domain.
Then\\
(i) the eigenvalues of $\mathcal{K}$ in $\Omega$   coincide with zeros of the function
\begin{align*}
&F(z):=\mathcal{K}_{00}-z-(\widehat{\mathcal{G}}(z)\kappa,\kappa^{*}), \\
&\kappa=(\mathcal{K}_{10},\mathcal{K}_{20},\dots),\,\,\kappa^*=(\mathcal{K}^*_{10},\mathcal{K}^*_{20},\dots);
\notag\end{align*}
(ii) if $F(z)$ has a unique root $z_*$ in $\Omega$ (and $F'(z_*)\ne 0$), then 
 the resolvent $\mathcal{G}(z)=(\mathcal{K}-z)^{-1}$  can be represented in the form
\begin{align}\label{p_sp.1}
&\mathcal{G}_{ij}(z)=\frac{\eta_i\bar\eta^*_j}{F'(z_*)(z_*-z)}+R_{ij}(z),\quad \eta_i=\eta_i(z_*),\quad \eta_{j}^*=\eta_{j}^*(z_*),\\
&\eta_i(z)=\delta_{0i}-(1-\delta_{0i})(\hat{\mathcal{G}}(z)\kappa)_i,\quad \eta_{j}^*(z)=\delta_{0j}-(1-\delta_{0j})(\hat{\mathcal{G}}^*(z)\kappa^*)_j,
\label{p_sp.1a}\end{align}
where $R_{ij}(z)$ is an analytic in $\Omega$ matrix-function  whose norm satisfies the bound
\begin{align}\label{p_sp.2}
\|R\|\le \sup_{z\in \Omega} \|\widehat{\mathcal{G}}\|+\sup_{z\in \Omega}\Big|\frac{d}{dz}\frac{z-z_*}{F(z)}\Big|(1+\|G\|\cdot \|\kappa\|)(1+\|G\|\cdot\|\kappa^*\|)\\
+\sup_{z\in \Omega} \Big|\frac{z-z_*}{F(z)}\Big|(1+\|\widehat{\mathcal{G}}\|) (1+\|G\|\cdot \|\kappa\|)(1+\|G\|\cdot \|\kappa^*\|).
\notag\end{align}
 \end{proposition}

\textit{Proof.} The assertion (i)  follows  from the standard Schur inversion formula valid for any $z:\,F(z)\not =0$:
\begin{align}\label{p_sp.2a}
&\mathcal{G}_{ij}(z)=\widehat{\mathcal G}_{ij}(z)+\frac{\eta_{i}(z)\eta^*_{j}(z)}{F(z)},
\end{align}
where we  set $\widehat{\mathcal G}_{ij}(z)=0$ if $i=0$ or $j=0$.

To prove  the assertion (ii), we write 
\[R_{ij}(z)=\hat{\mathcal{G}}_{ij}(z)+\frac{\eta_{i}(z)\eta^*_{j}(z)}{F(z)}-\frac{\eta_{i}(z_*)\eta^*_{j}(z_*)}{F'(z_*)(z-z_*)},\]
and use
\begin{align*}
\Big|\frac{\eta_{i}(z)\eta^*_{j}(z)(z-z_*)}{F(z)}-\frac{\eta_{i}(z_*)\eta^*_{j}(z_*)}{F'(z_*)}\Big|\le |z-z_*|\cdot \sup_{z\in \Omega}\Big|\dfrac{d}{dz}\Big(\frac{\eta_{i}(z)\eta^*_{j}(z)(z-z_*)}{F(z)}\Big)\Big|
\end{align*}
and 
\[\hat{\mathcal{G}}'(z)=\hat{\mathcal{G}}^2(z)\] 
to obtain the bound (\ref{p_sp.2}). $\quad\quad \square$
\medskip

Analysis of spectral properties of $\mathcal{K}$ is based on the analysis of $K$ of (\ref{K}).  Recall the definitions (\ref{A}), (\ref{A_1})
and choose $W,n$-independent $\delta>0$, which is small enough to provide that the domain
$\omega_\delta=\{x\in\mathbb{R}: |\mathcal{F}_0(x)|>1-\delta\}$ contains two non intersecting sub domains  $\omega_\delta^{+}$, $\omega_\delta^{-}$, 
such that each of $\omega_\delta^{+}$, $\omega_\delta^{-}$ contains one of the points $x=a_+$ and $x=a_-$ of maximum $\mathcal{F}_0(x)$
(easier speaking,  $\omega_\delta^{+}$, $\omega_\delta^{-}$ are two non-intersecting neighbourhood of points $a_+$ and $a_-$
(see Lemma \ref{l:f_a,b})).
Set also $\omega_{1,\delta}=\{x\in\mathbb{R}: |\mathcal{F}_1(x)|>1-\delta\}$.

To choose  a convenient basis in $L_2[\mathbb{R}]$, take $c_*:\Re c_*>0$, set
\begin{align*}
 &\alpha=\sqrt{\dfrac{c_*}{2}}\Big(1+\frac{c_*}{2W^2}\Big)^{1/2}=:\alpha_1+i\alpha_2,
\end{align*}
 and consider the system of the
 functions
\begin{align}\label{pA.1}
&\psi_0(x)=e^{-\alpha W x^2}\sqrt[4]{\alpha W/\pi};\\
&\psi_k(x)=h_k^{-1/2}e^{-\alpha Wx^2} e^{2\alpha_1W x^2}\Big(\frac{d}{dx}\Big)^ke^{-2\alpha_1W x^2}
=e^{-\alpha Wx^2}p_k(x);\notag\\
&h_k=k!(4\alpha_1W)^{k-1/2}\sqrt{2\pi},\quad k=1,2,\ldots
\notag\end{align}
It is easy to see that $p_k$ is the $k$th polynomials, orthogonal with the weight $e^{-2\alpha_1Wx^2}$
(it is the $k$th Hermite polynomial of $x\sqrt{2\alpha_1W}$ with a proper normalization).

Now let $\{\psi_{k}\}_{k=0}^\infty$ be  (\ref{pA.1}) with $c_*=c_+$ of (\ref{c_pm}). Consider the set  $\{\psi_{k,\delta}^+\}$ obtained by the
Gramm-Schmidt orthonormalization procedure of 
\[
\psi_k^+(x)=\psi_k(x-a_+)
\]
on $\omega_\delta^{+}$. 
Since $\psi^+_{k,\delta}(x)=O(e^{-cW})$ for $x\not\in\omega^+_\delta$, one can obtain easily
\begin{align}
&\psi^+_{k,\delta}(x)=\psi_{k}^+(x)+O(e^{-cW}),\quad k\ll W.
\notag\end{align}
By the same way we construct $\{\psi'_{k}(x)\}_{k=0}^\infty$ and $\{\psi'_{k,\delta}(x)\}_{k=0}^\infty$ on $\omega_{1,\delta}$ (with $b_s$ instead of $a_+$), 
and $\{\psi_{k}^-(x)\}_{k=0}^\infty$ and $\{\psi_{k,\delta}^-(x)\}_{k=0}^\infty$ on $\omega_\delta^{-}$ (with $c_*=c_-$ and $a_-$ instead of $a_+$). Everywhere
below we take
\begin{align*}
m=[\log^2 W]
\end{align*}
and consider two vector systems 
\begin{align}\label{Psi^+}
\{\Psi^+_{\bar k}(a,b)\}_{|k|\le m}=\{\psi^+_{k_1,\delta}(a)\psi'_{k_2,\delta}(b)\}_{|k|\le m}, \\ 
\{\Psi^-_{\bar k}(a,b)\}_{|k|\le m}=\{\psi^-_{k_1,\delta}(a)\psi'_{k_2,\delta}(b)\}_{|k|\le m}.
\notag\end{align}
Denote  $P^{+}$ and $P^-$ the projections on the subspaces spanned on the systems
$\{\Psi_{k,\delta}^+\}_{|k|\le m}$ and $\{\Psi_{k,\delta}^-\}_{|k|\le m}$ respectively.
 Evidently these projection operators are orthogonal to each other. 
Set
\begin{align}\label{P_i}
P=P^++P^-,\quad\mathcal{L}_1 =P\mathcal{H}, \quad\mathcal{L}_2=(1-P)\mathcal{H},\quad \mathcal{H}=\mathcal{L}_1\oplus\mathcal{L} _2,
\end{align}
where $\mathcal{H}=L_2(\mathbb{R}^2)$.

Note also that for any $u$ supported in some domain $\Omega$ and any $C>0$ 
\begin{equation}\label{razm_K}
(Ku)(a,b)=O(e^{-cW^2})\,\,\hbox{for}\, (a,b): \hbox{dist}\{(a,b),\Omega\}\ge C>0.
\end{equation}
Now consider the operator $K$ as a block operator with respect to the decomposition (\ref{P_i}).  It has the form
\begin{align}\label{K_21.0}
&K^{(11)}=K^+\oplus K^-+O(e^{-cW^2}),\\ 
& K^+:=P^{+}KP^{+},\quad K^{-}=P^{-}KP^{-},\notag\\
&K^{(12)}=P^{+}K(I^+-P^+)\oplus P^{-}K(I^{-}-P^{-})+O(e^{-cW}), \notag\\
& K^{(21)}=(I^{+}-P^{+})KP^{+}\oplus(I^{-}-P^{-})KP^{-}+O(e^{-cW}),
\notag\end{align}
where $I^{+}$ and $I^-$ are the operator of the multiplication by $1_{\omega_{\delta}^{+}}1_{\omega_{1,\delta}\vphantom{\omega^+_\delta}}$ and
 $1_{\omega_{\delta}^{-}}1_{\omega_{1,\delta}\vphantom{\omega^+_\delta}}$ respectively. Indeed,
 (\ref{razm_K})  and
\[
(A\psi_k^+)(x)=O(e^{-cW}) \,\,\hbox{for}\, |x-a_+|\ge C>0,
\]
yield   $P^+KP^- f=O(e^{-cW^2})$, $P_{-}K(I^+-P^+)f=O(e^{-cW})$, etc.
 
 Let $\hat K$  be $K$ without the  line and the column, corresponding to $\Psi^+_{\bar 0}$, 
 and  $\hat K^{+}$, is  defined similarly.
 Denote also
 \begin{align}\label{kappa}
&\kappa^+=K\Psi_{\bar 0}^+-(K\Psi_{\bar 0}^+,\Psi_{\bar 0}^+)\Psi_{\bar 0}^+, 
&\kappa^+_* =K^*\Psi_{\bar 0}^+-(K\Psi_{\bar 0}^+,\Psi_{\bar 0}^+)\Psi_{\bar 0}^+,
\notag\end{align}
 and set
 \begin{equation} \label{alp}
 \alpha_+=\sqrt{\dfrac{c_+}{2}}\Big(1+\dfrac{c_+}{2W^2}\Big)^{1/2}=: \alpha_1+i\alpha_2.
\end{equation} 
\begin{equation}\label{lam_0}
\lambda_{0,+}=\Big(1+\dfrac{2\alpha_+}{W}+\dfrac{c_+}{W^2}\Big)^{-1/2}.
\end{equation}

  \begin{theorem}\label{t:2} Given an operator $K$ of the form (\ref{K}), we have
  \begin{equation*}
 \Big| |\lambda_0(K)|-|\lambda_{0,+}|^2\Big|\le CW^{-3/2}.
  \end{equation*}
Moreover, for  any $z$ satisfying conditions
\begin{align}\label{z}
1-\dfrac{5\alpha_1}{2W}< |z|\le 1+\dfrac{C_0+5\alpha_1/2}{W},\quad |z-|\lambda_{0,+}|^2|\ge c/W.
 \end{align}
 with $C_0$ of (\ref{norm_S}) we have
 \begin{align}\label{t2.1}
& \|(\hat K^{(11)}-z)^{-1}\|\le CW, \\
& \|\hat K^{(12)}\|\le Cm/W,\quad\|\hat K^{(21)}\|\le Cm^{3/2}/W^{3/2},
\label{t2.2} \\
& \|K^{(22)}\|\le 1-Cm^{1/3}/W.
\label{t2.1a}
\end{align}
In addition,
\begin{align}\label{b_kappa}
&\|\kappa^+\|\le C/W^{3/2},
\quad \|\kappa^+_*\|\le C/W,\\
\label{00}
&(K\Psi_{\bar 0}^+,\Psi_{\bar 0}^+)=\lambda_{0,+}^2+O(W^{-3/2}),
\end{align}
and there is $0<q<1$  such that for all $|k|,|k'|\le m$
\begin{align}\label{repr_G}
(P^+\hat GP^+)_{\bar k,\bar k'}=&WG^{(ev)}_{\bar k,\bar k'}+W^{1/2}G^{(r)}_{\bar k,\bar k'},\\
|G^{(ev)}_{\bar k,\bar k'}|+ |G^{(r)}_{\bar k,\bar k'}|\le &Cq^{|\bar k-\bar k'|/2},\quad G^{(ev)}_{\bar k,\bar k'}=0,\,\hbox{if}\,\,
\bar k-\bar k'\not \in 2\mathbb{Z}^2.
\notag\end{align}
the vectors $\eta$ and $\eta^*$ defined as in (\ref{p_sp.1}) satisfy the conditions 
\begin{align}\label{t2.eta}
\eta=&\,\Psi_{\bar 0}^++W^{-1/2}\tilde\eta,\quad |\tilde\eta_{\bar k}|\le Cq^{|\bar k|/2},\\
\eta^*=&\,\eta^{*(ev)}+W^{-1/2}\tilde\eta^*,\quad |\eta^{*(ev)}_{\bar k}|+ |\tilde\eta^{*}_{\bar k}|\le Cq^{|\bar k|/2},
\,\,\,\eta^{*(ev)}_{\bar k}=0,\,\, \hbox{if}\,\,\, \bar k\not \in 2\mathbb{Z}^2.
\notag\end{align}
\end{theorem}   
Defer  the proof of Theorem \ref{t:2} to the next section and continue the analysis of $\mathcal{K}$.
 Write $\mathcal{K}$ as  
\begin{align}\label{t1.2}
 \mathcal{K}=\left(\begin{array}{cc} K^{(11)}S& K^{(12)}S\\
  K^{(21)}S& K^{(22)}S\end{array}\right),
\end{align}
  where
  \[ 
 K^{(\alpha\alpha')}S= \left(\begin{array}{cc} K^{(\alpha\alpha')}S_{11}& K^{(\alpha\alpha')}S_{12}\\
 K^{(\alpha\alpha')}S_{21}& K^{(\alpha\alpha')}S_{22}\end{array}\right) .
  \] 
Since all vectors in $\{\Psi_{\bar k}^+ \}_{|k|\le m}$ and $\{\Psi_{\bar k}^- \}_{|k|\le m}$ possess
the property
\begin{align*}
|\Psi_{\bar k}^+(a,b)|\le e^{-c\log^2W}, \quad \mathrm{if }\quad |a-a_+|+|b-b_s|\ge CW^{-1/2}\log W,\\
|\Psi_{\bar k}^-(a,b)|\le e^{-c\log^2W}, \quad\mathrm{ if} \quad |a-a_-|+|b-b_s|\ge CW^{-1/2}\log W
\end{align*}
for sufficiently big $C>0$, we have 
\[|S-S^+|\,\Psi^+_{\bar k}(a,b)=O(W^{-3/2}\log W),\quad|S-S^-|\,\Psi^-_{\bar k}(a,b)=O(W^{-3/2}\log W),\]
 where $S^+$ and $S^-$ have the form (\ref{T_W}) with $L$ replaced by $L^+$ and $L^-$ of (\ref{L_pm}) respectively.
Hence
 \[   
  K^{(11)}S=\left(\begin{array}{cc} K^+S^++O(W^{-3/2}\log W)&O(e^{-cW^2})\\
O(e^{-cW^2})& K^-S^-+O(W^{-3/2}\log W)\end{array}\right).
  \] 
It is useful to rewrite $\mathcal{K}$ in a more convenient form.  Write 
 \[S^+=V\Lambda V^{-1},\quad \Lambda=\left(\begin{array}{cc}\lambda^+_1&0\\0&\lambda^+_2\end{array}\right),
\quad V=\left(\begin{array}{cc}v_{11}&v_{12}\\v_{21}&v_{22}\end{array}\right)\]
 where   $\lambda^+_1$ and $\lambda^+_2$ are eigenvalues of $S^+$:
 \begin{equation}\label{evS^+}
 \lambda^+_1=1+\frac{L^+}{2W^2}+\sqrt{\frac{L^+}{W^2}+\frac{(L^+)^2}{4W^4}},\quad \lambda^+_2=1/\lambda^+_1,
 \end{equation} 
 and $V$ is a $2\times 2$ matrix diagonalizing $S^+$. It is easy to check that the eigenvectors of $S^+$ have the form
 $(v_{11},v_{21})= (-\sqrt{L^+},1)+O(W^{-1})$, $(v_{12},v_{22})= (\sqrt{L^+},1)+O(W^{-1})$, hence 
 \begin{equation}\label{norm_V}
 \|V\|\le C,\quad \|V^{-1}\|\le C.
 \end{equation}
Introduce the  matrix 
\[ \breve{V}=\left(\begin{array}{ccc} V&0&0\\0&I&0\\0&0&I\end{array}\right), 
\]
where the first block corresponds to $K^+S^+$, the second one to $K^-S^-$, and the third one to $ K^{(22)}S$.
Now set
\begin{align}\label{K_V}
\mathcal{K}_V=\breve V^{-1}\mathcal{K}\breve V=\left(\begin{array}{ccc} K^+\Lambda+O\Big(\frac{\log W}{W^{3/2}}\Big)&O(e^{-cW^2})&K^{+(12)}V^{-1}S\\
O(e^{-cW^2})&K^-S^-+O\Big(\frac{\log W}{W^{3/2}}\Big)&K^{-(12)}S\\K^{+(21)}SV&K^{-(21)}S&K^{(22)}S\end{array}\right).
\end{align}
Note that 
\begin{equation}\label{K^-S}
K^-S=K^-S^-+O(W^{-3/2}\log W)=\left(
\begin{array}{cc}
K^-&0\\
-K^-/W&K^-
\end{array}\right)
+O(W^{-3/2}\log W). 
\end{equation}
Then
 (\ref{main1}) and (\ref{main2}) can be written as
\begin{align}\label{main*}
\bar g_n(E)=&\,\frac{W}{n}\sum_{j=0}^{n-1}\Big( \big(\mathcal{K}^j_V\breve{\mathcal{B}}\mathcal{K}^{n-1-j}_Ve_V,e_V^*\big)_2
\mathcal{F},\bar{\mathcal{F}}\Big),
\quad e_V=V^{-1}e_2,\quad e_V^*=V^{*}e_L,\\
1=&\,W\Big( \big(\mathcal{K}^{n-1}_Ve_V,e_V^*\big)_2
\mathcal{F},\bar{\mathcal{F}}\Big).
\label{main**}\end{align}
Now we can formulate the main result of the section
\begin{theorem}\label{t:1} Given an operator $\mathcal{K}_V$ defined in (\ref{K_V}) we have
\begin{align}\label{t1.0}
\lambda_0(\mathcal{K}_V)=1,\quad |\lambda_1(\mathcal{K}_V)|\le 1-c/W.
\end{align}
Moreover,  the resolvent $\mathcal{G}(z)=(\mathcal{K}_V-z)^{-1}$ can be written as
\begin{align}\label{t1.R}
\mathcal{G}(z)=&\frac{P_\eta}{1-z}+R(z),\quad \|R\|\le CW,\quad\mathrm{if}\quad z\in\Omega_0=\{|z|\ge 1-c/2W\},
\end{align}
where $P_\eta$ is a rank one operator of the form
\begin{align}\label{t1.P}
P_\eta&= \eta_V
\otimes\eta_V^*,\quad
\eta_V=\eta\otimes e_1+O(W^{-1}),\quad \eta^*_V=\eta^*\otimes e_1+O(W^{-1}), 
\end{align}
with $\eta$, $\eta^*$  of (\ref{t2.eta}).
\end{theorem}
\textit{Proof.} Prove first that
\begin{align}\label{t1.1}
|\lambda_0(\mathcal{K}_V)-1|\le C\log^2 W/W^{3/2}.
\end{align}
In order to apply Proposition \ref{p:sp}, we want to prove first
that  for $\hat{\mathcal{G}}(z)=(\hat{\mathcal{K}}_V-z)^{-1}$ 
\begin{align}\label{t1.7}
 \|\hat{\mathcal{G}}(z)\|\le CmW,\quad\hbox{if}\quad z\in\Omega=\{1-\alpha_1/3W\le  |z|\le 1+C_0/W\}
  \end{align}
with $\alpha_1$, $C_0$ of (\ref{z}).

According  to (\ref{K_V}) and the formula for the inverse of the block matrix, it is easy to see that to prove (\ref{t1.7})
  it  suffices to check 
   \begin{align}\label{t1.4}
& \| (\hat K^{+}\lambda_1^+-z)^{-1}\|\le CW,\quad \,\, \| ( K^{+}\lambda_2^+-z)^{-1}\|\le CW, \quad  \| ( K^{-}S^--z)^{-1}\|\le CW,\\
&\| K^{+(12)}V^{-1}S\|\le Cm/W,\quad \| K^{-(12)}S\|\le Cm/W,\notag\\
&\| K^{+(21)}SV\|\le C(m/W)^{3/2},\quad\,\,  \| K^{-(21)}S\|\le C(m/W)^{3/2},\notag\\
&\|(K^{(22)}S-z)^{-1}\|\le CW/m^{1/3}
\notag \end{align}
for $z\in\Omega$.  
Rewrite
\[
(\hat K^{+}\lambda_1^+-z)^{-1}=(\lambda_1^+)^{-1}(\hat K^+-z/\lambda_1^+)^{-1},\quad
(\hat K^{+}\lambda_2^+-z)^{-1}= \lambda_1^+(\hat K^+-z/\lambda_2^+)^{-1}.
\] 
Note that according to (\ref{L_pm}) and (\ref{evS^+})
\begin{equation*}
\lambda^+_1=\lambda_{0,+}^{-2}.
\end{equation*} 
 Now using (\ref{lam_0}) it is easy to check  that if $z\in\Omega$ of (\ref{t1.7}), then both $z/\lambda_1^+$ and $z/\lambda_2^+$ satisfy (\ref{z}), and therefore 
 we get the first line of (\ref{t1.4}).
The second and the third lines follow  from (\ref{t2.1}) -- (\ref{t2.2}), (\ref{K^-S}) and  (\ref{norm_S}), (\ref{norm_V}).
Moreover, (\ref{norm_S}) and the representation (\ref{t1.2}) combined with (\ref{t2.2}) -- (\ref{t2.1a}) yield 
\[
\|K^{(22)}S\|\le 1-\dfrac{Cm^{1/3}}{W}.
\]
Hence
 \begin{align}\label{t1.5}
  \|(K^{(22)}S-z)^{-1}\|\le \dfrac{1}{|z|}\sum_{s=0}^{\infty}\dfrac{\|K^{(22)}S\|^s}{|z|^s}\le C_1W/m^{1/3},
  \end{align}
which finishes the proof of (\ref{t1.4}), thus (\ref{t1.7}).

Now we can apply Proposition \ref{p:sp}  to $\mathcal{K}_V$ and consider 
\begin{multline*}
F(z)=\lambda_1^+K^+_{\bar 0\bar 0}-z-(\hat{ \mathcal{G}}(z)\kappa_V,\kappa^*_V)+O\Big( \frac{\log W}{W^{3/2}}\Big)\\
=\lambda_{0,+}^2\lambda^+_1-z
-(\hat{ \mathcal{G}}(z)\kappa_V,\kappa^*_V)+O\Big( \frac{\log W}{W^{3/2}}\Big),
\end{multline*}
where $\kappa_V$ and $\kappa^*_V$ are the column and the line of $\mathcal{K}_V$ that correspond to the vector $\Psi_{\bar 0}\otimes e_1$. 
To get the second equality here we used (\ref{00}).
By (\ref{b_kappa}) and (\ref{norm_V})
\begin{align}\label{b_kappa_V}
\|\kappa_V^*\|\le CW^{-1},\quad \|\kappa_V\|\le CW^{-3/2}.
\end{align}
Define also (recall $\lambda_{0,+}^2\lambda_1^+=1$)
\[
F_0(z)=\lambda_{0,+}^2\lambda^+_1-z=1-z,\quad \sigma=\{z:|z-1|\le C_*W^{-3/2}\log^2 W\}
\] 
with sufficiently big $C_*$. Then (\ref{b_kappa_V}) and (\ref{t1.7}) yield
\[
|(\hat{ \mathcal{G}}(z)\kappa_V,\kappa^*_V)|\le Cm/W^{3/2}\Rightarrow|F_0(z)|>|F(z)-F_0(z)|, \quad z\in \partial\sigma,
\]
and hence the Rouche theorem implies that $F(z)$, $F_0(z)$ have the same number of zeros (one) in $\sigma$, which gives (\ref{t1.1}). 
Taking $z\in \sigma(z_0)=\{z\in\mathbb{C}: |z-z_0|\le C_*\log^2 W/W^{3/2}\}$ for any $|z_0-\lambda_0(\mathcal{K}_V)|>2C_*/W$ satisfying (\ref{t1.R}),
 by the same argument one can obtain that $F(z)$ has the same number of zeros as $F_0(z)$  in $\sigma(z_0)$ (i.e. zero, since $\lambda_{0,+}^2\lambda^+_2$, 
 $|\lambda_{0,+}|^2$ do not satisfy (\ref{t1.R})), which
 implies the second bound of (\ref{t1.0}) with $c=\alpha_1/3$. The representation (\ref{t1.R}) follows from (\ref{p_sp.1})-(\ref{p_sp.2}) if we take into account
(\ref{t2.eta}),  (\ref{t1.7}), (\ref{b_kappa_V}) and the fact that $$\kappa_V=\lambda_{1}^+\kappa\otimes e_1+O(W^{-3/2}).$$

Let us prove now the first relation in (\ref{t1.0}). The Cauchy formula for the resolvent yields
 \[\mathcal{K}_V^{n-1}=\frac{1}{2\pi i}\oint_{L}z^{n-1}\mathcal{G}(z)dz\]
 for any closed contour $L$ which contains all  eigenvalues of $\mathcal{K}$. Let us choose $L$ as a union of  two circles:
\[
L_1=\{z:|z|=1-c/2W\},\quad L_0=\{z:|z-1|\le c/3W\}
\]
with some sufficiently small but $n,W$-independent $c$. 
Hence we get
\begin{align*}
W\Big( \big(\mathcal{K}_V^{n-1}e_V,e_V^*\big)_2
\mathcal{F},\bar{\mathcal{F}}\Big)=\frac{W}{2\pi i}\Big(\oint_{L_0}+\oint_{L_1}\Big)z^{n-1}\Big( \big(\mathcal{G}(z)e_V,e_V^*\big)_2\mathcal{F},\bar{\mathcal{F}}\Big)\\
=\frac{W}{2\pi i}\oint_{L_0}z^{n-1}dz\Big( \big(\mathcal{G}(z)e_V,e_V^*\big)_2\mathcal{F},\bar{\mathcal{F}}\Big)+O(W^2e^{-cn/W})=: I_1^0+O(mW^2e^{-cn/W}),
\end{align*}
where the bound for the remainder follows from (\ref{norm_V}) and the bound on the norm of the resolvent on the contour $L_1$ obtained from (\ref{t1.R}):
\begin{align}\label{bG}
\|\mathcal{G}(z)\|\le CmW,\quad z\in L.
\end{align}
But using the representation (\ref{t1.R}) -- (\ref{t1.P}), definition of $e_V$, $e_V^*$  in (\ref{main*}),  and the Cauchy theorem we have
\begin{align*}
I_1^0=W\lambda_0^{n-1}(\mathcal{K}_V)(\tilde \eta_V,\bar{\mathcal{F}})(\tilde \eta_V^*,\bar{\mathcal{F}}),
\end{align*}
where 
\begin{equation}\label{eta_til}
\tilde \eta_V=(V^{-1})_{12}\eta\otimes e_1+O(W^{-1}),\quad \tilde \eta_V^*=(V^*e_L)_1\eta^*\otimes e_1+O(W^{-1})
\end{equation}
with $\eta$, $\eta^*$ of (\ref{t2.eta}). 
Hence, on the basis of (\ref{main**}) we conclude that
\begin{align*}
1=W\lambda_0^{n-1}(\mathcal{K}_V)(\tilde \eta_V,\bar{\mathcal{F}})(\tilde \eta_V^*,\bar{\mathcal{F}})+O(e^{-cn/2W}).
\end{align*}
Using that $\lambda_0(\mathcal{K}_V)$ and $(\tilde \eta_V,\bar{\mathcal{F}})(\tilde \eta_V^*,\bar{\mathcal{F}})$ do not depend on $n$, and $n$ in the above formula can be taken
arbitrary large, we conclude that 
\begin{align}\label{1=}
\lambda_0(\mathcal{K}_V)=1,\quad W(\tilde \eta_V,\bar{\mathcal{F}})(\tilde \eta_V^*,\bar{\mathcal{F}})=1,
\end{align}
which gives the first equality in (\ref{t1.0}).
$\square$

\textit{Proof of Theorem \ref{t:0}.}      
Set
\begin{align}\label{B^c}
\mathcal{B}^{\circ}=\mathcal{B}-(E/2+i\sqrt{4-E^2}/2)I
\end{align}

Evidently
\begin{align*}
\sum_{j=0}^{n-1}\mathcal{K}^j_V\mathcal{B}^{\circ}\mathcal{K}^{n-1-j}_V=\frac{1}{(2\pi i)^2}\oint_{L}dz_1\oint_{L'}dz_2\frac{z_1^n-z_2^n}{z_1-z_2}\mathcal{G}(z_1)\mathcal{B}^{\circ}\mathcal{G}(z_2),
\end{align*}
where the contour $L$ was chosen above and the contour $L'=L_0'\cup L_1'$ is chosen similarly, but on the distance $d/W$ from $L$ with some sufficiently small 
fixed $d$. Then  by (\ref{main1}) and the above formula we obtain
\begin{align*}
\bar g_n(E)=n^{-1}(I_1+I_2+I_3+I_4),
\end{align*}
where $I_1$ corresponds to the integral over $z_1\in L_0, z_2\in L_0'$, $I_2$ corresponds to the integral over $z_1\in L_0, z_2\in L_1'$, $I_3$
corresponds to the integral over $z_1\in L_1, z_2\in L_0'$, and $I_4$ corresponds to the integral over $z_1\in L_1, z_2\in L_1'$. 
The bound for the resolvent (\ref{bG}), and the estimates
\begin{align*}
\qquad\|[\mathcal{B}^{\circ},\mathcal{G}(z_2)]\|&=\|\mathcal{G}(z_2)[\mathcal{B}^{\circ},\mathcal{K}_V]\mathcal{G}(z_2)\|\le Cm^2W\\
\Rightarrow\|\mathcal{B}^{\circ}\mathcal{G}(z_2)\mathcal{F}\|&\le\|\mathcal{G}(z_2)\mathcal{B}^{\circ}\mathcal{F}\|+\|[\mathcal{B}^{\circ},\mathcal{G}(z_2)]\mathcal{F}\|\\
&\le CmW\|\mathcal{B}^{\circ}\mathcal{F}\|+Cm^2W\|\mathcal{F}\|\le Cm^2W
\end{align*}
yield for some absolute $p>0$
\[|I_4|=\Big |\frac{W}{(2\pi i)^2}\oint_{L_1}dz_1\oint_{L'_1}dz_2\frac{z_1^n-z_2^n}{z_1-z_2}\Big( \big(\mathcal{G}(z_1)\mathcal{B}^{\circ}\mathcal{G}(z_2)e_V,e_V^*\big)_2\mathcal{F},\bar{\mathcal{F}}\Big)\Big|\le Cm^pW^4e^{-nc/W}.\]
Let us  prove the bound for $I_2$.  We write first
\begin{align*}
I_2&=\frac{W}{(2\pi i)^2} \oint_{L_0}dz_1\oint_{L'_1}dz_2\frac{z_1^n}{z_1-z_2}\Big( \big(\mathcal{G}(z_1)\mathcal{B}^{\circ}\mathcal{G}(z_2)e_V,e_V^*\big)_2\mathcal{F},\bar{\mathcal{F}}\Big)+
O(m^pW^4e^{-nc/W})\\
&=\frac{W}{2\pi i} \oint_{L'_1}\frac{dz_2}{1-z_2}\Big( \big(P_\eta\mathcal{B}^{\circ}\mathcal{G}(z_2)e_V,e_V^*\big)_2\mathcal{F},\bar{\mathcal{F}}\Big)+
O(m^pW^4e^{-nc/W})\\
&=\frac{W}{2\pi i}  \lim_{R\to\infty}\Big(\oint_{|z|=R}-\oint_{L'_0}\Big)\frac{dz_2}{1-z_2}\Big( \big(P_\eta\mathcal{B}^{\circ}\mathcal{G}(z_2)e_V,e_V^*\big)_2\mathcal{F},\bar{\mathcal{F}}\Big)+
O(m^pW^4e^{-nc/W})\\
&=-\frac{W}{2\pi i} \oint_{L'_0}\frac{dz_2}{1-z_2}\Big( \big(P_\eta\mathcal{B}^{\circ}\mathcal{G}(z_2)e_V,e_V^*\big)_2\mathcal{F},\bar{\mathcal{F}}\Big)+
O(m^pW^4e^{-nc/W})\\
&= I_2'+O(m^pW^4e^{-nc/W}).
\end{align*}
To estimate $I_2'$, observe that  by (\ref{p_sp.2a}) 
\begin{align*} 
&\Big(\big(P_\eta\mathcal{B}^{\circ}\mathcal{G}(z)e_V,e_V^*\big)_2\mathcal{F},\bar{\mathcal{F}}\Big)=
C_*(V)(\mathcal{F},\eta)(\mathcal{B}^{\circ}G(z)\mathcal{F},\eta^*\Big)\\
=&C_*(V)(\mathcal{F},\eta)\Big(\frac{(\mathcal{F},\Psi_{\bar 0})-(\mathcal{F},\hat G^*\kappa^*)}{F(z)}\Big( (\Psi_{\bar 0}, \mathcal{B}^{\circ}\eta^*)-
(\hat G\kappa, \mathcal{B}^{\circ}\eta^*)\Big)
+(\mathcal{F}, \hat G^*\mathcal{B}^{\circ}\eta^*)\Big),
\end{align*}
where $C_*(V)$ is some constant  depending on the entries of $V$, $G=(K-z)^{-1}$, $\hat G=(\hat K-z)^{-1}$, $\kappa,\kappa^*,\eta,\eta^*$ are defined as in Proposition \ref{p:sp}.
We will prove that
\begin{align}\label{bd.1}
&|(\mathcal{F},\Psi_{\bar 0})|\le CW^{-1/2},\quad|(\mathcal{F},\hat G\kappa^*)|\le CW^{-1/2},\quad|(\mathcal{F},\eta)|\le CW^{-1/2},\\
&|(\Psi_{\bar 0}, \mathcal{B}^{\circ}\eta^*)|\le CW^{-1},\quad |(\hat G\kappa, \mathcal{B}^{\circ}\eta^*)|\le CW^{-1},\label{bd.2}\\
&|(\mathcal{F}, \hat G^*\mathcal{B}^{\circ}\eta^*)|\le CW^{-1/2}.
\label{bd.3}\end{align}
Since $|z_2-1|^{-1}=cW$ on the contour $L_0'$ and the length of $L_0'$ is $2\pi(cW)^{-1}$, this inequalities will give us
\[ |I_2'|\le C \Rightarrow I_2=O(1).\]

The first inequality of (\ref{bd.1}) can be obtained by the direct calculations. The second and third follow from the bounds
(\ref{t2.eta}) and (\ref{repr_G})
\[|\eta_{\bar k}|\le Cq^{|\bar k|/2},\quad |(\hat G\kappa^*)_{\bar k}|\le C q^{|\bar k|/2},\quad |\bar k|\le m.\]
 Observe that by (\ref{repr_G}) and (\ref{t2.eta})
\[(\mathcal{B}^{\circ}\eta^*)(a,b)=(b-b_s)\eta^{*(ev)}(a,b)+W^{-1/2}(b-b_s)\tilde\eta^{*}(a,b)+O(W^{-m/4}),
\]
where $\eta^{*(ev)}(a,b)$ contains the sum of $\Psi^+_{(2k_1,2k_2)}(a,b)$, and $\tilde\eta^{*}(a,b)$ contains the sum
of  $\Psi_{(k_1,k_2)}(a,b)$ (with any $\bar k$) with  exponentially decreasing coefficients. Then the structure of $\hat G$ (\ref{repr_G}) implies that
\begin{align*}
&(\hat G^*\mathcal{B}^{\circ}\eta^*)(a,b)=W(b-b_s)\nu^{(ev)}(a,b)+W^{1/2}(b-b_s)\tilde\nu(a,b)+O(W^{-m/4}),\\
&|\nu^{(ev)}_{\bar k}|+|\tilde\nu^{*}_{\bar k}| \le Cq^{|\bar k|/2} ,\quad(|\bar k|\le m)
\end{align*}
where $\nu^{(ev)}(a,b)$ still contains only $\Psi_{\bar k}$ with $\bar k\in2\mathbb{Z}^2$.
It is easy to see that by (\ref{Psi^+}) for $|k|\le m$ we have
\begin{align*}
\int \mathcal{F}(a,b)&(b-b_s)\Psi^+_{(k_1,2k_2+1)}(a,b)dadb\\=&(2\alpha W)^{1/2}\int e^{f_a(a)}p_{k_1}((2\alpha W)^{1/2}(a-a_+))e^{-\alpha W(a-a_+)^2}da\\
&\times\int e^{f_b(b)}(b-b_s)p_{2k_2+1}((2\alpha W)^{1/2}(b-b_s))e^{-\alpha W(b-b_s)^2}db\\
=&(2\alpha W)^{-1}\int  p_{k_1}(a)e^{-a^2/2}(1+O(a^2/W))da\\
&\times\int  b\,p_{2k_2+1}(b)e^{-b^2/2}(1+O(b^2/W^2))db+O(e^{-c\log^2 W})=O(|k|W^{-1}),
\end{align*}
where $\{p_k\}_{k=0}^\infty$ are  normalized Hermit polynomials (with a weight $e^{-x^2}$). 

The same argument applied to $\Psi^+_{(k_1,2k_2)}$ yields
\begin{align*}
\int \mathcal{F}(a,b)&(b-b_s)\Psi^+_{(k_1,2k_2)}(a,b)dadb=O(|k|W^{-5/2}),
\end{align*}
thus we obtain (\ref{bd.3}). Bounds (\ref{bd.2}) can be obtained similarly.
 The same argument yields also
\[I_3=O(1).\]
Now using the identity
\begin{align*}
\frac{1}{(2\pi i)^2} \oint_{L_0}\oint_{L'_0}\frac{z_1^n-z_2^n}{z_1-z_2}\frac{dz_1dz_2}{(1-z_1)(1-z_2)}=n,
\end{align*}
 the representation (\ref{p_sp.1}), and the Cauchy theorem, we get
\begin{align*}
I_1&=\frac{W}{(2\pi i)^2} \oint_{L_0}dz_1\oint_{L'_0}dz_2\frac{z_1^n-z_2^n}{z_1-z_2}\Big( \big(\mathcal{G}(z_1)\mathcal{B}^{\circ}\mathcal{G}(z_2)e_V,e_V^*\big)_2\mathcal{F},\bar{\mathcal{F}}\Big)\\
&=Wn(\mathcal{B}^{\circ}\eta_V,\eta^*_V)(\tilde \eta_V,\bar{\mathcal{F}})(\tilde \eta^*_V,\bar{\mathcal{F}}),
\end{align*}
where $\tilde \eta_V$, $\tilde \eta^*_V$ are defined in (\ref{eta_til}). 
Thus, according to (\ref{1=}), we have
\begin{align*}
\bar g_n(E)=(\mathcal{B}^{\circ}\eta_V,\eta^*_V)+(E/2+i\sqrt{4-E^2}/2)+O(n^{-1}).
\end{align*}
Hence,  to finish the proof of the theorem, it suffices to show that
\begin{align}\label{Be,e}
|(\mathcal{B}^{\circ}\eta,\eta^*)|\le CW^{-1}\Rightarrow|(\mathcal{B}^{\circ}\eta_V,\eta^*_V)|\le CW^{-1}.
\end{align}
But  the first line of (\ref{t2.eta}),  the definition (\ref{pA.1}) of $\psi_k$, and the definition (\ref{B^c}) of $\mathcal{B}^{\circ}$ yield
\[\eta=\Psi_{\bar 0}+O(W^{-1/2})\Rightarrow \mathcal{B}^{\circ}\eta=cW^{-1/2}\Psi_{\bar{01}}+O(W^{-1}).\]
Combining  this with the second line of (\ref{t2.eta}) we obtain (\ref{Be,e}).
$\square$

\section{Proof of Theorem \ref{t:2}}\label{s:5}
 Let us first introduce  the ``model" operator
\begin{align*}
&A_*^{(c_*)}(x,y)=\mathcal{F}_*(x)B(x,y)\mathcal{F}_*(y),\quad \mathcal{F}_*(x)=e^{-c_*x^2/2}, \quad \Re c_*>0.
\end{align*}
The main properties of $A_*$ are given by the following lemma, proved in \cite{SS:16} (see Lemma 3.1):
\begin{lemma}\label{l:A_*} Given an orthonormal system
 $\{\psi_k\}_{k\ge 0}$  defined in (\ref{pA.1}), we have
\begin{align*}
A_*^{(c_*)}\psi_0=\lambda_0^{(c_*)}\psi_0,\quad \lambda_0^{(c_*)}=\Big(1+\frac{2\alpha}{W}+\frac{c_*}{W^2}\Big)^{-1/2}.
\end{align*}
The matrix $A_{*jk}^{(c_*)}:=(A_*^{(c_*)}\psi_k,\psi_j)$ is  upper triangular, $(A_*^{(c_*)})_{jk}=0$, if $j$ and $k$ have
 different evenness, and
\begin{align}\label{pA.3}
&A_{*kk}^{(c_*)}=(\lambda_0^{(c_*)})^{2k+1},\quad
A_{*k,k+2}^{(c_*)}=-2i\alpha_2\frac{\sqrt{(k+1)(k+2)}}{W}\,\Big(1+O\Big(\frac{k+1}{W}\Big)\Big),\\
&|A_{*k,k+2p}^{(c_*)}|\le \frac{C^p(k+1)^p}{W^p}.
\label{pA.4}\end{align}
In addition, if  $\{\tilde\psi_k\}$ are defined by (\ref{pA.1}) with $c_*$ replaced by some $c_0>0$, and 
$\tilde P_l$ is a projection on  the space spanned on $\{\tilde\psi_r\}_{k=0}^l$, and $P_m$ is a similar projection for $\{\psi_k\}_{k=0}^m$,
then
\begin{align}\label{P_lm}
\|\tilde P_l(1-P_m)\|\le Cl^3/m.
\end{align}
\end{lemma}

Recall that $m=[\log^2 W]$, and thus $\psi_{k,\delta}^+(y)$ is $O(e^{-c\log^2W})$ for $|y-a_+|\ge 2CW^{-1/2}\log W$
  (for sufficiently big $C>0$ and $k\le m$). Therefore, we have
\begin{equation*}
A\psi_{k,\delta}^+(x)= O(e^{-c\log^2W}),  \quad |x-a_+|\ge CW^{-1/2}\log W,
\end{equation*}
where $A$ is defined in (\ref{A}). In addition,
$A\psi_{k,\delta}^+(x)$ can be written in the form ($k<m$)
\[A\psi_{k,\delta}^+(x)=\int _{|y-a_+|\le C\log W/W^{1/2}}(A_*^+(x-a_+,y-a_+)+\widetilde A_+(x,y))\psi_{k,\delta}^+(y)dy+O(e^{-c\log^2W}). 
\]
Here and below we denote
\begin{equation*}
A_*^\pm:=A_*^{(c_\pm)},\quad A^\pm_m=P_m^{\pm} A P^{\pm,}_m\quad \widetilde A_+(x,y)=A(x,y)-A_*^+(x-a_+,y-a_+),
\end{equation*}
where the projections $P_m^+$ and $P^{-}_m$ are defined like in (\ref{P_lm}) for $\{\psi_{k,\delta}^+\}_{k=0}^m$ and $\{\psi_{k,\delta}^-\}_{k=0}^m$.

Expanding $\mathcal{F}_0$ for $|x-a_+|\le CW^{-1/2}\log W$, $|y-a_+|\le CW^{-1/2}\log W$, we get in this neighbourhood
\begin{align*}
\widetilde A(x,y)=A(x,y)\, O(\log^3 W/W^{3/2} ).
\end{align*}
 Thus, for $k\le m$
\begin{align}\label{G.3b}
A\psi_k^+(x)=(A^+_*\psi_k)(x-a_+)+O(\log^3 W/W^{3/2}),
\end{align}
and similarly for $A_1$ of (\ref{A_1})
\begin{align}\label{G.3a}
A_1\psi_k'(x)=(A^+_*\psi_k)(x-b_s)+O(\log^3 W/W^{3/2}).
\end{align}
\begin{remark}\label{r:a}
Applying the Taylor expansions  up to the $m$-th order to the functions $\mathcal{F}_0(x)$ and $\mathcal{F}_0(y)$ 
 one can prove
that for $j,k=0,\ldots, m$
\begin{align}\label{r_a.1}
|A^\pm_{m,jk}|\le 
\left\{
\begin{array}{ll}
(Cm_{j,k}/W)^{|j-k|/2},& j-k\ge 2;\\
(Cm_{j,k}/W)^{3/2},& j-k=1;\\
(Cm_{j,k}/W)^{|j-k|/2},& k-j\ge  3;\\
(Cm_{j,k}/W)^{3/2},& k-j=1,2.
\end{array}
\right.,
\end{align}
where $m_{j,k}=\max \{j,k\}$. 
In addition, 
\begin{align}\label{r_a.12}
&\|(A^{(12)}_{\pm})_j\|\le \left\{
\begin{array}{ll}
(Cm/W)^{|m-j|/2},& m-j\ge 2;\\
Cm/W,& m-j=1,\\
\end{array}\right.\\ \notag
&\|(A^{(21)}_{\pm})_j\|\le \left\{
\begin{array}{ll}
(Cm/W)^{|m-j|/2},& m-j\ge 3;\\
(Cm/W)^{3/2},& m-j=1,2,\\
\end{array}\right.
\end{align}
where $(A^{(12)}_{\pm})_j$ and $(A^{(21)}_{\pm})_j$ are the $j$-th row and column of $A^{(12)}_\pm=P_\pm A (I_\pm-P_\pm)$ and $A^{(21)}_\pm=(I_\pm-P_\pm)A P_{\pm}$ respectively ($I_\pm$ here are operators of multiplication by $1_{\omega_\delta^\pm}$).
Indeed,  it is well known that 
the Hermite functions $\{\psi_k(x)\}_{k=0}^\infty$ satisfy the recursion relation
\[
 x\psi_k(x)=\sqrt{\frac{k+1}{4\alpha_1W}}\psi_{k+1}(x)+\sqrt{\frac{k}{4\alpha_1W}}\psi_{k-1}(x).
 \]
Hence,  the operator $\widehat L$ of multiplication by $x-a_+$ has a three diagonal form in the  basis $\{\psi_k^+\}$,  and 
$\widehat L^l$ has $2l+1$ non empty diagonals. The recursion relations combined with (\ref{pA.4}) yield (\ref{r_a.1}).
To prove (\ref{r_a.12}) we have to use also 
\begin{align*}
&\|(A^{(12)}_{\pm})_j\|^2=\|A^*\psi_{j,\delta}^{\pm}\|^2-\|(A_m^{\pm})^*\psi_{j,\delta}^{\pm}\|^2+O(e^{-cW}),\\
&\|(A^{(21)}_{\pm})_j\|^2=\|A \psi_{j,\delta}^{\pm}\|^2-\|A_m^{\pm}\psi_{j,\delta}^{\pm}\|^2+O(e^{-cW}).
\end{align*}
Similar bounds hold for $A_1$, thus for $K=A\otimes A_1$ (probably with multiplication by $m^p$ with some absolute $p>0$).
\end{remark}
\medskip
\textit{Proof of (\ref{t2.2}).} By (\ref{K_21.0}), to prove the bound for $\|\hat K^{(12)}\|$, we need to prove bounds for $\|P^+\hat K(1-P^+)\|$
and $\|P^-\hat K(1-P^-)\|$. Let $P_m$ and $P_{1,m}$ be the projections on $\{\psi_k(x-a_+)\}_{0\le k\le m}$ and $\{\psi_k(x-b_s)\}_{0\le k\le m}$. Then
\begin{align*}
&\|P^+\hat K(1-P^+)\|=\|(P_m\otimes P_{1,m})(\hat A\otimes\hat A_1)(1-P_m\otimes P_{1,m})\|
\\
&=\|(P_m\otimes P_{1,m})(\hat A\otimes\hat A_1)((1-P_m)\otimes 1+1\otimes(1- P_{1,m})-(1-P_m)\otimes(1- P_{1,m}))\|\\
&\le||\hat A^{(12)}\|+||\hat A^{(12)}_1\|+||\hat A^{(12)}\|\cdot \|\hat A^{(12)}_1\|.
\end{align*}
Hence it suffices to prove that
\begin{align*}
||\hat A^{(12)}\|\le Cm/W,\quad ||\hat A^{(12)}_1\|\le Cm/W,
\end{align*}
which follows from (\ref{r_a.12}).
By the same way  one can estimate $\|(I^+-P^+)KP^+\|$,  $\|P^-K(I^- -P^-)\|$, $\|(I^- -P^-)KP^-\|$ and prove (\ref{b_kappa}).

$\square$

\medskip

\textit{Proof of (\ref{t2.1}), (\ref{repr_G}), and (\ref{t2.eta})}.  Since $\hat K^{(11)}=\hat K^+\oplus \hat K^-+ O(e^{-cW^2})$, it suffices to prove the
bound for $\|(\hat K^+-z)^{-1}\|$ and $\|(\hat K^--z)^{-1}\|$ . The bounds are very similar, hence we prove only the first one.

Consider
the diagonal matrix with the entries
\[D_{\bar k\bar k}=A^+_{*k_1k_1}A^+_{*k_2k_2}-z,\quad 0< |\bar k|\le m.
\]
According to (\ref{z}), (\ref{lam_0}) and (\ref{pA.3}) we get for $|k|>0$
\begin{align}\label{d_bound}
&|D_{\bar k\bar k}|>\Big|1-\dfrac{\alpha_+(2k_1+1)}{W}-\dfrac{\alpha_+(2k_2+1)}{W}+O(W^{-2})-z\Big|\\ \notag
&\ge \Big||z|-\Big|1-\dfrac{\alpha_+(2k_1+1)}{W}-\dfrac{\alpha_+(2k_2+1)}{W}+O(W^{-2})\Big|\Big|\\ \notag
&\ge \dfrac{\alpha_1(2k_1+2k_2+3/2)}{W}-\dfrac{5\alpha_1}{2W}+O(W^{-2})= \dfrac{\alpha_1(2k_1+2k_2-1-\varepsilon)}{W}.
\end{align}
Hence 
\[
\|D^{-1}\|\le CW.
\]
Set $R=( \hat K^+-D-z)D^{-1}$. Let $Q$ be the matrix which contains  $O(1)$-order (or higher order) entries of
$R$  while the others entries are replaced by zeros. It follows from (\ref{pA.4}), (\ref{G.3b}) -- (\ref{G.3a}) that  
\begin{align}\label{Q} 
Q_{\bar k \bar k'}\not=0\quad \mathrm{iff}\quad \bar k'-\bar k=2e_1\vee \bar k'-\bar k=2e_2 \quad (e_1=(1,0),\, e_2=(0,1)).
\end{align}
Using the notations we can rewrite
\begin{align}\label{r_tild}
(\hat K^+-z)^{-1}=&D^{-1}(I+R)^{-1}=D^{-1} (1+Q)^{-1}(I+\widetilde R)^{-1}\\
=&D^{-1} (1+Q)^{-1}(I-\widetilde R(I+\widetilde R)^{-1}),
\notag\end{align}
where $\widetilde R=(R-Q)(I+Q)^{-1}$. 

Moreover, there exists an absolute constant $l_{\alpha}$ such that for $|k|>l_\alpha$
\begin{align} \label{norm_Q}
&|Q_{\bar k,\bar k+2e_1}|+|Q_{\bar k,\bar k+2e_2}|\le 
\frac{|A^+_{*k_1k_1}A^+_{*k_2,k_2+2}|}{\alpha_1(2k_1+2k_2+3-\varepsilon)}+\frac{|A^+_{*k_1,k_1+2}A^+_{*k_2,k_2}|}{\alpha_1(2k_1+2k_2+3-\varepsilon)}\\
\le&\frac{\alpha_2}{\alpha_1}\frac{\sqrt{(k_1+1)(k_1+2)}+\sqrt{(k_2+1)(k_2+2)}}{(k_1+k_2+(3-\varepsilon)/2)}\le (\alpha_2/\alpha_1)^{1/2}=q<1.
\notag\end{align}
Here we used (\ref{pA.3}), (\ref{d_bound}) and the fact $\alpha_2<\alpha_1$ (see (\ref{alp})  and use $\arg c_\pm\in (-\pi/2,\pi/2)$).

Write $Q$ as a block matrix   
\begin{align*}& Q^{(11)}=\{Q_{\bar k,\bar k'}\}_{|k|\le l_{\alpha},|k'|\le l_{\alpha}},\quad
Q^{(12)}=\{Q_{\bar k,\bar k'}\}_{|k|\le l_{\alpha},|k'|> l_{\alpha}},\\
& Q^{(21)}=\{Q_{\bar k,\bar k'}\}_{|k|> l_{\alpha},|k'|\le l_{\alpha}}\quad
Q^{(22)}=\{Q_{\bar k,\bar k'}\}_{|k|> l_{\alpha},|k'|> l_{\alpha}}
\end{align*}

 Then by (\ref{Q}) $Q^{(21)}=0$, and by (\ref{norm_Q}) $||Q^{(22)}||\le q$. Moreover, (\ref{Q}) implies that for $s_0=[l_{\alpha}/2]+1$
 \[Q^{s_0}=\left(\begin{array}{cc}0&X\\0&(Q^{(22)})^{s_0}\end{array}\right)\Rightarrow
Q^{s_0+p}=\left(\begin{array}{cc}0&X(Q^{(22)})^{p}\\0&(Q^{(22)})^{s_0+p}\end{array}\right),\quad p>0, 
 \]
where $X$ is some fixed matrix. 

Writing  the Neumann series $(1+Q)^{-1}=\sum_s (-1)^sQ^s$ we obtain that in view of (\ref{Q})
\begin{align}\label{Q_elem}
|(1+Q)^{-1}_{\bar k,\bar k'}|\le Cq^{|\bar k-\bar k'|/2},
\end{align}
and, in addition,  
\begin{align}\label{Q_ev}
(1+Q)^{-1}_{\bar k\bar k'}=0,\,\, \hbox{if}\,\, \bar k-\bar k'\not\in 2\mathbb{Z}^2.
\end{align}
Note that below $0<q<1$ can be different in different formulas. 

Besides, it is easy to check using (\ref{r_a.1}) and (\ref{d_bound}) that 
$$\|R-Q\|\le m^{p}W^{-3/2},\quad |(R-Q)_{\bar k\bar k'}|\le (Cm/W)^{|\bar k-\bar k'|/2}.$$
Here and below we denote by $p, p_1, p_2$ etc. some absolute exponents which could be different in different formulas.
Hence
 \begin{align*}
&|\widetilde R_{\bar k\bar k'}|=\Big|\sum_{|\bar k''|\le m}(R-Q)_{\bar k\bar k''}
(1+Q)^{-1}_{\bar k'',\bar k'}\Big|\le C/W^{1/2}q^{|\bar k-\bar k'|/2}.
\end{align*}
The last relation implies
\begin{align}
 |(1+\widetilde R)^{-1}_{\bar k\bar k'}|\le Cq^{|\bar k-\bar k'|/2}.
\label{b_R}\end{align} 
To prove this, let us consider any fixed $\bar k$ and $\bar k'$ and  use the standard trick from the spectral theory
(see e.g. \cite{PS:11}, Ch. 13.3). Assume that $|\bar k-\bar k'|=k_1-k_1'$. Then denote $ D_q$ the diagonal matrix
such that $( D_q)_{\bar k''\bar k'''}=\delta_{\bar k''\bar k'''}q^{k''_1/2}$. Then
 \begin{align*}
& ||D_q^{-1}\widetilde RD_q||\le Cm^{p}/W^{1/2}\\
\Rightarrow & |(1+\widetilde R)^{-1}_{\bar k\bar k'}|=
 |(D_q(1+D_q^{-1}\widetilde RD_q)^{-1}D_q^{-1})_{\bar k\bar k'}|\\
& \le q^{(k_1-k_1')/2}||(1+D_q\widetilde RD_q^{-1})^{-1}||.
\end{align*} 
If $|\bar k-\bar k'|=-(k_1-k_1')$ we use $D_q^{-1}$ instead of $D_q$. And if $|\bar k-\bar k'|=\pm(k_2-k_2')$ we use
$( D_q)_{\bar k''\bar k'''}=\delta_{\bar k''\bar k'''}q^{\pm k''_2/2}$.
The last line of (\ref{r_tild}) combined with (\ref{b_R}) proves (\ref{t2.1}) and the representation similar to (\ref{repr_G}) for $(\hat K^+-z)^{-1}$
with 
\[
G^{(ev)}=W^{-1}D^{-1} (1+Q)^{-1},\quad G^{(r)}=-W^{-1/2}D^{-1}(1+Q)^{-1}\widetilde R (I+\widetilde R)^{-1}.
\]
Conditions of the second line of (\ref{repr_G}) hold because of (\ref{d_bound}) and  (\ref{Q_elem}) --  (\ref{b_R}).

Now let us use the standard linear algebra formula
\begin{align*}
P^+\hat GP^+=(\hat K^+-z-\widetilde R_1)^{-1},\quad \widetilde R_1=\hat K^{(12)}_+(K^{(22)}_+-z)^{-1}\hat K^{(21)}_+,
\end{align*}
where
\[
K^{(12)}_+=P_+K(I_+-P_+),\quad K^{(12)}_+=(I_+-P_+)KP_+,\quad K^{(22)}_+=(I_+-P_+)K(I_+-P_+).
\]
Assume for the moment that (\ref{t2.1a}) is known already, which gives  (see (\ref{t1.5})) $$\|(K^{(22)}-z)^{-1}\|\le CW/m^{1/3}.$$ 
Together with (\ref{t2.1}) -- (\ref{t2.2}) this implies
\[
\|(K^{(22)}_+-z)^{-1}\|\le mW.
\]
Then 
we obtain for $|\bar k|, |\bar k'|\le m$
\begin{align*}
| (\widetilde R_{1})_{\bar k,\bar k'}|&=|(\hat K^{(12)}_+(K^{(22)}_+-z)^{-1}\hat K^{(21)}_+)_{\bar k,\bar k'}|= |\sum\limits_{\bar k'', \bar k'''}
\hat K^{(12)}_{+,\bar k,\bar k''}(K^{(22)}_+-z)^{-1}_{\bar k'',\bar k'''}\hat K^{(21)}_{+,\bar k''',\bar k'}|\\
&\le CmW\sum\limits_{\bar k'', \bar k'''}|\hat K^{(12)}_{+,\bar k,\bar k''}|\cdot |\hat K^{(21)}_{+,\bar k''',\bar k'}|,
\end{align*}
which together with  (\ref{r_a.12}) implies
\[| (\widetilde R_{1})_{\bar k,\bar k'}|\le m^p(Cm/ W)^{|\bar k-\bar k'|/2}.
\]
Now, using the trick applied above to prove (\ref{b_R}) and the formula
\[(\hat K^+-z-\widetilde R_1)^{-1}=(\hat K^+-z)^{-1}-(\hat K^+-z)^{-1}\widetilde R_1(I-\widetilde R_1(\hat K^+-z)^{-1})^{-1}\]
 one can to obtain (\ref{repr_G})  from the representation for $(\hat K^+-z)^{-1}$.

Representation (\ref{t2.eta}) follows from (\ref{repr_G}), the definition of $\eta$, $\eta^*$ (\ref{p_sp.1a}), and (\ref{r_a.1}).

$\square$

\medskip

\textit{Proof of  (\ref{t2.1a}).}
Let us split  the integration domain $\mathbb{R}^2$ into  three sub domains, according to the value of the functions $\mathcal{F}$, $\mathcal{F}_1$.
One of the possible splitting is 
\begin{align*}
&\Lambda_1=\{(a,b):|\mathcal{F}(a)|,|\mathcal{F}_1(b)|\ge 1- \delta/2\},\\
&\Lambda_3=\{(a,b):1-\delta>|\mathcal{F}(a)|,|\mathcal{F}_1(b)|\},\notag\\
&\Lambda_2=\mathbb{R}^2\setminus(\Lambda_1\cup\Lambda_3).\notag
\end{align*}
Write
\[u=u_1+u_2+u_3,\]
where $u_i=u1_{(a,b)\in\Lambda_i}$. Since $\max_{(a,b)\in\Lambda_2\cup\Lambda_3}|\mathcal{F}(a)\mathcal{F}_1(b)|=1-\delta/2$,
we have 
\begin{align}\notag
||Ku||^2\le ||(\mathcal{F}\mathcal{F}_1)^2u||^2\le& ||u_1||^2+(1-\delta/2)^2||u_2+u_3||^2\\
=&1-(1-(1-\delta/2)^2)||u_2+u_3||^2\notag\\
\Rightarrow&||u_2+u_3||^2\le C_0(1-||Ku||^2).
\label{u_2}\end{align}
Here we  used that the operator with the kernel $B(a_1,a_2)$ defined by (\ref{A_1}) satisfies the bound $||B||\le 1$. This is true, 
if the integration
with respect to $a_1,a_2$ is over the real line. If the integration contour is deformed (see Remark \ref{r:b_L} ), then
\[||B||\le \sup_a|\cos^{-1/2}2\phi(a)|.\]
But if the condition (\ref{b_L.1}) is satisfied, then the inequality (\ref{u_2}) is still true (may be with some different $C_0$).

 Moreover,
\begin{align}\label{u_3}
\Re(K (u_1+u_2),Ku_{3})&=\Re(K u_1,Ku_{3})+\Re(K u_2,Ku_{3})\\
&=O(e^{-cW^2})+\Re(Ku_2, Ku_{3})\le O(e^{-cW^2})
+\frac{1}{2}\|u_2+u_3\|^2,
\notag
\end{align}
since $\|K\|\le 1$.
Denote 
\begin{align}\label{u^pm_0}
u_0:=u_1+u_2,\quad  u_0^{+}=u_0 1_{\omega_\delta^{+}},\quad
u_0^{-}=u_0 1_{\omega_\delta^{-}}.
\end{align}  
\begin{lemma}\label{l:K_22a}
For  $u_0^{+}$ and $u_0^{-}$ defined in (\ref{u^pm_0}) we have
\begin{align}\label{K_22a.1}
&\|Ku_0^{+}\|^2\le (1-Cm^{1/3}/W)\|u_0^{+}\|^2,\quad \|Ku_0^{-}\|^2\le (1-Cm^{1/3}/W)\|u_0^{-}\|^2.
\end{align}
\end{lemma}
Assume for the moment that the lemma is proved and finish the proof of (\ref{t2.1a}).

We have by (\ref{u_2}),  (\ref{u_3}), and the lemma
\begin{align*}
||Ku||^2=&\|K(u_0^{+}+u_0^{-}+u_3)\|^2\notag
\\=&||Ku_0^+||^2+||Ku_0^-||^2+2\Re(K u_0,Ku_{3})+||Ku_{3}||^2+O(e^{-cW})\notag\\
\le &(1-Cm^{1/3}/W)(\|u_0^{+}\|^2+\|u_0^{-}\|^2)+
2||u_2+u_3||^2+O(e^{-cW})\notag\\
\le &1-Cm^{1/3}/W+2C_0(1-||Ku||^2)\notag\\
\Rightarrow&(1-||Ku||^2)(1+2C_0)\ge Cm^{1/3}/W\notag\\
\Rightarrow &||Ku||^2\le 1-C_1m^{1/3}/W.
\end{align*}
Here we used that
\[
\|Ku_3\|^2\le \|u_3\|^2\le \|u_2\|^2+\|u_3\|^2=\|u_2+u_3\|^2.
\]
$\square$

\medskip

\textit{ Proof of Lemma \ref{l:K_22a}.}  

Choose $c_0>0$ sufficiently small to provide
\[\Re f(x)\ge \dfrac{c_0}{2}(x-a_+)^2,\quad x>0,\]
and denote
\[
\mathcal{F}_0=e^{-c_0(x-a_+)^2/2},\quad A_0=\mathcal{F}_0B\mathcal{F}_0.
\]
 Consider the basis $\{\widetilde\psi_k\}_{k\ge 0}$ defined by (\ref{pA.1}) for $c_0$. 
Define the operator kernel $A_{0,+}(a_1,a_2):=A_0(a_1,a_2)$ and
similarly define $A_{0,1}(b_1,b_2)$ (with $b_s$ instead of $a_+$). Since $c_0$ is real, $A_{0,+}$ and $A_{0,1}$ are diagonal in the basis
 $\{\widetilde\psi_k\}_{k\ge 0}$ .
 Moreover,  the commutator $[\mathcal{F},B]$  admits the bound
\begin{align*}
\|\,[\mathcal{F},B]\,\|\le&\sup_{x}W\int|\mathcal{F}(x)-\mathcal{F}(y)|e^{-W^2(x-y)^2}dy
\le& C W\int|x-y|e^{-W^2(x-y)^2}dy\le C_*/W,
\end{align*}
and by the same argument
\[\|[\mathcal{F}_0,B]\,\|\le C_*/W.\]
Thus, denoting $\Lambda^+$ the projection on $\omega_\delta^+$ we get   
\begin{align*}
\Lambda^+ A^*\Lambda^+ A \Lambda^+&=\Lambda^+\mathcal{F}^*B\mathcal{F}^*\Lambda^+\mathcal{F}B\mathcal{F}\Lambda^+=\Lambda^+ B\mathcal{F}^*\mathcal{F}^*\Lambda^+\mathcal{F}\mathcal{F}B\Lambda^+
+\Lambda^+ [\mathcal{F}^*,B]
\mathcal{F}^*\Lambda^+\mathcal{F}\mathcal{F}B\Lambda^+\\
&+\Lambda^+B \mathcal{F}^*\mathcal{F}^*\Lambda^+\mathcal{F}[B,\mathcal{F}]\Lambda^+
+\Lambda^+ [\mathcal{F}^*,B]\mathcal{F}^*\Lambda^+\mathcal{F}[B,\mathcal{F}]\Lambda^+\\ 
&\le \Lambda^+ B\mathcal{F}^*\mathcal{F}^*\Lambda^+\mathcal{F}\mathcal{F}B\Lambda^++3C_*W^{-1}\\
&\le \Lambda^+ BF_0^4B\Lambda^++3C_*W^{-1}\le \Lambda^+\mathcal{F}_0B\mathcal{F}_0^2B \mathcal{F}_0\Lambda^++6C_*W^{-1}\\
&= \Lambda^+ A_0A_0\Lambda^++6C_*W^{-1},
\end{align*}
and similarly
\[\Lambda_1  A_1^*\Lambda_1 A_1\Lambda_1 \le \Lambda_1  A_0^2\Lambda_1 +6C_*W^{-1},\]
where $\Lambda_1$ is the projection on $\omega_{1,\delta}$.
Set
\[
K_0=A_{0,+}\otimes A_{0,1}, \quad  \Lambda=\Lambda^+\otimes \Lambda_1.
\]
Then, taking into account that $\|A\|,\|A_1\|\le 1$, we obtain
\begin{align}\label{K<K_0}
\Lambda K^*\Lambda K\Lambda &=\Lambda^+ A^*\Lambda^+ A\Lambda^+ \otimes \Lambda_1 A_1^*\Lambda_1 A_1\Lambda_1 
\le\Lambda^+ A_0^2 \Lambda^+ \otimes \Lambda_1 A_1^*\Lambda_1 A_1\Lambda_1 +6C_*W^{-1}\\&\le
\Lambda^+ A_0^2 \Lambda^+ \otimes\Lambda_1  A_{0,1}^2\Lambda_1 +12C_*W^{-1}=\Lambda K_0^2\Lambda+12C_*W^{-1}.
\notag\end{align}
 Let $\tilde P_l$ and $\tilde P_{1,l}$ be the projection operator on  $\{\widetilde\psi_k(x-a_+)\}_{0\le k\le l}$ and $\{\widetilde\psi_k(x-b_s)\}_{0\le k\le l}$ 
  respectively, while $P_m$ and $P_{1,m}$ be the projections on $\{\psi_k(x-a_+)\}_{0\le k\le m}$ and $\{\psi_k(x-b_s)\}_{0\le k\le m}$. By
  (\ref{P_lm})
\begin{align}\label{P_lm.1}
&\|\tilde P_l\otimes\tilde P_{1,l}\Big(1-P_m \otimes P_{1,m}\big)\|\\
&=\big\|\tilde P_l\otimes\tilde P_{1,l}\big((1-P_m) \otimes I+I \otimes(1- P_{1,m})
-(1-P_m )\otimes (1-P_{1,m})\big)\big\|\notag\\
&\le\|\tilde P_l(1-P_m )\|+\|\tilde P_{1,l}(1-P_{1,m})\|+\|\tilde P_l(1-P_m )\|\cdot \|\tilde P_{1,l}(1-P_{1,m})\|\le\dfrac{Cl^3}{m}\le \frac{1}{2},
\notag\end{align}
if $l=m^{1/3}/C_1$ with sufficiently big $C_1$.
Use now the following proposition
\begin{proposition}\label{p:22}
Let $\mathcal{A}\le I$ and $\mathcal{A}_0\le I$ be positive operators, $\Lambda$, $P$ and $\tilde P$ be projection operators such that
\begin{align}
&\notag\Lambda\mathcal{A}\Lambda\le \Lambda\mathcal{A}_0\Lambda+\delta,\quad\|[\Lambda,P]\|\le \delta_1,\quad \|[\Lambda,\tilde P]\|\le \delta_1,\\
&[\tilde P,\mathcal{A}_0]=0,\quad (1-\tilde P)\mathcal{A}_0(1-\tilde P)\le 1-\Delta,\notag\\
&\|\tilde P(1-P)\|\le\frac{1}{2}.
\label{P_lm.2}\end{align}
Then
\begin{align}\label{p_22.1}
\Lambda(1-P)\mathcal{A}(1-P)\Lambda\le 1-\Delta/2+\delta+4\delta_1.
\end{align}
\end{proposition}

Apply the proposition to $\mathcal{A}= K^*\Lambda K$, $\mathcal{A}_0=K_0$, $P=P^+=P_m\otimes P_{1,m}$,
$\tilde P=\tilde P_l\otimes\tilde P_{1,l}$. Then $\Delta=l\sqrt{2c_0}/2W$ since $\tilde P$ is a correspondent spectral projection
of $K_0$, $\delta=8C_*W^{-1}$ (by (\ref{K<K_0}), $\delta_1=O(e^{-cW})$, and (\ref{P_lm.2}) is valid in view (\ref{P_lm.1}).  Then
(\ref{p_22.1}) yields (\ref{K_22a.1}).

$\square$

\textit{Proof of Proposition \ref{p:22}.}

Take  any $u=(1-P)\Lambda v$, $\|v\|=1$. Then
\begin{align*}
(\mathcal{A}u,u)&\le ((1-P)\Lambda\mathcal{A}\Lambda(1-P)v,v)+2\delta_1\le (\mathcal{A}_0u,u)+\delta+4\delta_1\\
&= (\tilde P\mathcal{A}_0\tilde P u,u)+((1-\tilde P)\mathcal{A}_0(1-\tilde P) u,u)+\delta+4\delta_1\\
&\le\|\tilde Pu\|^2+(1-\Delta)(\|u\|^2-\|\tilde Pu\|^2)+\delta+4\delta_1\\
&\le(1-\Delta)+\Delta\|\tilde Pu\|^2+\delta+2(2\delta_1+\delta_1^2)\le(1-\Delta/2)+\delta+4\delta_1,
\end{align*}
since
\begin{align*}
&|((1-\tilde P)\mathcal{A}_0(1-\tilde P) u,u)|=|((1-\tilde P)^2\mathcal{A}_0(1-\tilde P)^2 u,u)|\le \|(1-\tilde P)u\|^2=\|u\|^2-\|\tilde Pu\|^2;\\
&\|\tilde Pu\|^2=\|\tilde P(1-P)\Lambda v\|^2\le \|\Lambda v\|^2/2\le 1/2.
\end{align*}

$\square$

\medskip

\textbf{Acknowledgement.}
We are  very grateful to Sasha Sodin, who drew our attention to the transfer matrix approach in application to 1d random band matrices, for many
fruitful discussions.

\end{document}